\documentclass[12pt,preprint]{aastex}  %REQUIRED FOR APJ SUBMISSION
\def\ang{\AA}

\def\gapprox{\lower.4ex\hbox{$\;\buildrel >\over{\scriptstyle\sim}\;$}}
\def\lapprox{\lower.4ex\hbox{$\;\buildrel <\over{\scriptstyle\sim}\;$}}

\def\ref#1{\par\noindent\hangindent1cm {#1}}

\slugcomment{Revised Version, 2007 Aug 27}
\shortauthors{ASCHWANDEN ET AL. 2007}
\shorttitle{SCALING LAWS OF SOLAR AND STELLAR FLARES}

\begin{document}
%{\sl Submitted to ApJ, Revised 2007 Apr 25; 2007 Aug 27}

\title{		Scaling Laws of Solar and Stellar Flares }

\author{        Markus J. Aschwanden and Robert A. Stern }

\affil{         Lockheed Martin Advanced Technology Center,
                Solar \& Astrophysics Laboratory,
                Org. ADBS, Bldg.252,
                3251 Hanover St.,
                Palo Alto, CA 94304, USA;
                e-mail: aschwanden@lmsal.com}

\and
\author{        Manuel G\"udel}

\affil{		 Institute for Astronomy,
		ETH Zentrum,
		CH-8092 Zurich, Switzerland}

\begin{abstract}
In this study we compile for the first time comprehensive data sets of solar 
and stellar flare parameters, including flare peak temperatures $T_p$, 
flare peak volume emission measures $EM_p$, and flare durations $\tau_f$ from
both solar and stellar data, as well as flare length scales $L$ from solar data.
Key results are that both the solar and stellar data are consistent with a common 
scaling law of $EM_p \propto T_p^{4.7}$, but the stellar flares exhibit  
$\approx 250$ times higher emission measures (at the same flare peak temperature). 
For solar flares we observe also systematic trends for the flare length scale
$L(T_p) \propto T_p^{0.9}$ and the flare duration $\tau_F(T_p) \propto T_p^{0.9}$
as a function of the flare peak temperature. Using the theoretical RTV scaling
law and the fractal volume scaling observed for solar flares,
i.e., $V(L) \propto L^{2.4}$, we predict a scaling law of
$EM_p \propto T_p^{4.3}$, which is consistent with observations, and a scaling law
for electron densities in flare loops, $n_p \propto T_p^2/L \propto T_p^{1.1}$. 
The predicted ranges of electron densities are 
$n_p \approx 10^{9-10}$ cm$^{-3}$ for solar nanoflares at $T_p=1$ MK, 
$n_p \approx 10^{10-11}$ cm$^{-3}$ for typical solar flares at $T_p=10$ MK, and
$n_p \approx 10^{11-12}$ cm$^{-3}$ for large stellar flares at $T_p=100$ MK. 
The RTV-predicted electron densities were also found to be consistent with
densities inferred from total emission measures, $n_p=\sqrt{EM_p/q_V V}$,
using volume filling factors of $q_V=0.03-0.08$ constrained by fractal
dimensions measured in solar flares. Solar and stellar flares are expected
to have similar electron densities for equal flare peak temperatures $T_p$,
but the higher emission measures of detected stellar flares most likely represents
a selection bias of larger flare volumes and higher volume filling factors, due
to low detector sensitivity at higher temperatures. 
Our results affect also the determination of radiative and conductive cooling 
times, thermal energies, and frequency distributions of solar and stellar flare 
energies.
\end{abstract}


\section{       INTRODUCTION 		                   	}

Scaling laws provide important diagnostics and predictions for specific physical 
models  of nonlinear processes such as self-organized criticality, turbulence, 
diffusion, plasma heating, and particle accleration. These models have been
widely applied in plasma physics, astrophysics, geophysics, and the biological 
sciences. Here we investigate scaling laws of physical parameters in solar and 
stellar flares, which should allow us to decide whether solar and stellar flare 
data are consistent with the same physical flare process. 

The scaling of solar and stellar flare data has been pioneered by Stern (1992), 
Feldman et al.~(1995b), and Shibata \& Yokoyama (1999; 2002), who showed evidence
for a nonlinear scaling between the flare volume emission measure $EM_p$ and
the flare peak temperature $T_p$. These parameters have been measured in
solar flares with instruments like {\sl Skylab}, {\sl GOES}, {\sl Yohkoh/SXT}
(Soft X-ray Telescope), and {\sl RHESSI (Ramaty High Energy Solar Spectroscopic 
Imager)}, and in stellar flares with {\sl ASCA}, {\sl BeppoSAX}, {\sl Einstein},
{\sl EUVE}, {\sl EXOSAT}, {\sl Ginga}, {\sl HEAO}, {\sl ROSAT}, {\sl Chandra}, 
and {\sl XMM-Newton}.
Compilations of solar flare parameters have been presented in Aschwanden (1999),
while stellar flare parameters were compiled in a recent review by G\"udel (2004).
In this paper we present for the first time this host of mostly new 
measurements ``on the same page'' and investigate commonalities and differences
between the scaling of solar and stellar flares. 

In Section 2 we present the statistical correlations found in stellar flare data,
while the corresponding counterparts of solar flare data are shown in Section 3.
In Section 4 we present theoretical modeling of the data, using the well-known
RTV law, the generalization with gravitational stratification and spatially
non-uniform heating, the fractal flare volume scaling, and volume filling factor. 
In Section 5 we discuss the differences between solar and stellar scaling laws, 
the consistency between two different electron density measurement methods, and 
a previously derived ``universal scaling law'' for solar and stellar flares. 
Section 6 summarizes our conclusions. 

\section{	STELLAR FLARE OBSERVATIONS			}

\subsection{	Definitions of Physical Parameters		}

A stellar coronal flare is usually detected from light curves in extreme ultra-violet (EUV)
or soft X-ray wavelengths, from which a (background-subtracted) peak count rate 
$c_p$ [cts s$^{-1}$] at the flare peak time $t=t_p$ can be measured.
The count rate $c(t)$ for optically-thin emission (as it is the case in EUV and
soft X-rays) is generally defined by the temperature integral of the total
(volume-integrated)
differential emission measure distribution $dEM(T)/dT$ [cm$^{-3}$] and the 
instrumental response function $R(T)$ (in units of [cts s$^{-1}$ cm$^3$]), 
\begin{equation}
        4 \pi d^2 \ c(t) = \int  {dEM(T) \over dT} \ R(T)\ dT \ ,
\end{equation}
where the factor $(4\pi d^2)$ comes from the total emission over the full celestial
sphere at a stellar distance $d$ (in parsecs). 
The differential emission measure distribution (DEM) of flares shows usually a 
single peak at the flare peak temperature $T_p$, so that the emission measure peak 
at the flare peak time, $EM_p = dEM(t=t_p,T)/dT \approx dEM(t=t_p, T=T_p)$, 
can be approximated with a single 
temperature (which corresponds to an emission measure-weighted average value),
\begin{equation}
        4\pi d^2 \ c_p = 4\pi d^2 \ c(t=t_p) \approx EM_p \ R(T_p) \ .
\end{equation}
The total (volume-integrated) emission measure $EM_p$ at the flare peak is 
defined as the squared electron density $n$ integrated over the source volume $V$,
\begin{equation}
        EM_p = \int n^2 dV \approx n_p^2 V \ ,
\end{equation}
where the right-hand approximation implies that $n_p^2=n^2(t=t_p,T=T_p)$ 
is the squared electron density at the flare peak time averaged
over the volume $V$ of the flare plasma, assuming a unity filling factor. 

Integrating the count rate $c(t)$ over the flare duration $\tau_f$ yields the
total counts $C$, which in the case of a single-peaked DEM can also be approximated 
(with Eq.~2) as
\begin{equation}
        4\pi d^2 \ C = 4\pi d^2 \int c(t) dt \approx 4\pi d^2 \ c_p \ \tau_f = EM_p\ R(T_p)\ \tau_f \ .
\end{equation}

The radiative loss rate for optically thin plasmas is a function of the squared density
and the radiative loss function $\Lambda (T)$,
\begin{equation}
	{dE_R \over dV\ dt} = n_e n_i \Lambda(T) \approx n_e^2 \Lambda(T) \ ,
\end{equation}
(in the coronal approximation of fully ionized plasma, i.e., $n_e \approx n_i$)
where the radiative loss function has a typically value of $\Lambda(T) \approx 
10^{-23...-22}$ [erg cm$^3$ s$^{-1}$] in the temperature range of $T\approx 10^{6...8}$ K.
From this we can define a peak luminosity $L_X$ in soft X-rays by integrating over
the volume and temperature range,
\begin{equation}
	L_X = V \int n^2(t=t_p, T) \Lambda(T) dT \approx EM_p \ \Lambda(T_p) \ .
\end{equation}
The total radiated energy $E_X$ integrated over the flare duration is then
\begin{equation}
	E_X = \int \int \int n^2(t,T) \Lambda(T) \ dV\ dT\ dt
	 \approx EM_p \ \Lambda(T_p) \ \tau_f \ .
\end{equation}
This yields a convenient conversion from observed total counts $4\pi d^2 \ C$ (Eq.~4)
into total radiated energy $E_X$ (Eq.~7),
\begin{equation}
	E_X = {\Lambda (T_p) \over R(T_p)} 4 \pi d^2 \ C = f(T_p) \ 4 \pi d^2 \ C \ , 
\end{equation}
which involves a temperature-dependent conversion factor $f(T_p)=\Lambda (T_p) / R(T_p)$.

For comparison we calculate also the total thermal energy $E_T$ of the flare volume at the flare peak time $t=t_p$,
\begin{equation}
        E_T = \int 3 n(t=t_p, T) k_B T(t=t_p) V(t=t_p) dT \approx 3 n_p k_B T_p V = 
		{3 k_B EM_p T_p \over n_p}
\end{equation}
where $n_p = n(t=t_p, T=T_p)$ represents the electron density at the flare peak time
$t=t_p$ and DEM peak temperature $T=T_p$.
The relation between the total thermal energy $E_T$ and the total radiated energy $E_X$
is then
\begin{equation}
	E_T \approx E_X {3 k_B T_p \over n_p(T_p) \ \Lambda(T_p) \ \tau_f(T_p)} \ ,
\end{equation}
where the peak electron density $n_p(T_p)$ and the flare duration $\tau_f(T_p)$ may
have a statistical dependence on the flare peak temperature $T_p$, and this way define
the temperature dependence in the correlation between the thermal energy $E_T$ and the
total radiated energy $E_X$.

\subsection{	Statistical Correlations		}

A recent compilation of stellar flare measurements is given in G\"udel (2004; Table 4
therein).  This database contains measurements of the total energy radiated in soft 
X-rays, $E^X$, the peak emission measure $EM_p$, flare peak temperature $T_p$, 
and flare duration $\tau_f$ from 68 different stellar flares.
Generally, only the peak emission measure $EM_p$ (derived from the peak count rate
$c_p$ [Eq.~2] and the known response function $R(T_p)$), the flare peak temperature
$T_p$ (if a EUV/soft X-ray spectrum is available), and the flare duration $\tau_f$
can be measured directly, which we will consider as independent variables in the
following, while all other quantities such as the energies ($E_X, E_T$) 
and densities $n_p$ are derived quantities, using relations as given in \S 2.1.

We show scatterplots of various parameters as a function of the flare peak temperature 
in Fig.~1. We perform linear regression fits of $y(x)$, $x(y)$ (shown with thin lines
in Fig.~1), with the {\sl ordinary least square bisector method} (shown with thick lines
in Fig.~1), and calculate the linear regression coefficients (indicated with
RC in Fig.~1). We find a highly significant correlation (with a regression 
coefficient of $RC=0.68$) for the flare peak emission measure $EM_p$ as a function 
of the flare peak temperature $T_p$ (Fig.~1 top left), i.e.,
\begin{equation}
        EM_p(T_p) = 10^{50.8} \left({T_p \over 10\ {\rm MK}}\right)^\alpha \ , 
			\quad \alpha = 4.5 \pm 0.4 \ .
\end{equation}
This value is essentially identical to the correlation given in G\"udel (2004),
where a powerlaw slope of $\alpha=4.30\pm0.35$ is quoted, determined with a similar
linear regression method. We test this general $EM_p-T_p$ scaling law in
evolutionary curves observed from 8 different stars in the next Section (\S 2.3). 

The other independently measured parameter is the flare duration $\tau_f$, for which we
find a marginally significant correlation ($RC=0.39$) with the flare temperature
$T_p$ (Fig.1 lower left), 
\begin{equation}
        \tau_f(T_p) = 10^{2.5} \left({T_p \over 10\ {\rm MK}}\right)^\beta \ , 
			\quad \beta = 1.8 \pm 0.2 \ .
\end{equation}

The other correlations can be understood as a consequence of the correlations
found between the independent parameters $T_p, EM_p$, and $\tau_f$. For instance for the
peak X-ray luminosity $L_X$ we find (Fig.~1 top right),
\begin{equation}
        L_X(T_p) = 10^{27.8} \left({T_p \over 10\ {\rm MK}}\right)^\gamma \ , 
			\quad \gamma = 4.7 \pm 0.4 \ .
\end{equation}
because the peak luminosity (Eq.~6) is proportional to the peak emission measure,
i.e., $L_X \propto EM_p$, if we neglect the weak temperature dependence of the
radiative loss function $\Lambda (T_p)$, and thus should have about the same
powerlaw slope, i.e., $\gamma=4.7 \approx \alpha=4.5$. 

For the total radiated X-ray energy $E_X$ we find a correlation of (Fig.~1 bottom 
right), 
\begin{equation}
        E_X(T_p) = 10^{30.7} \left({T_p \over 10\ {\rm MK}}\right)^\delta \ , 
			\quad \delta = 6.1 \pm 0.5 \ .
\end{equation}
which is expected to scale as $E_X(T_p) \propto EM_p(T_p) \tau_f(T_p)
\propto T_p^\delta$ (Eq.~7) with $\delta = \alpha + \beta = 4.5 + 1.8 = 6.3$,
which indeed agrees with the best fit, $\delta = 6.1 \pm 0.5$. 

We cannot calculate the thermal flare energy $E_T$ (Eq.~10) with current data,
since we need the knowledge of the flare peak density $n_p$, which is generally not 
independently measured from $EM_p$ and $T_p$. The values quoted in Table 4 of
G\"udel (2004) have a mean value of $n_p = 10^{11.5 \pm 0.6}$ cm$^{-3}$ and
show no temperature dependence. The flare peak density $n_p$ can be calculated 
with relation (Eq.~10),
if one uses some theoretical assumptions (e.g., RTV scaling law, see \S 4.1)
which predicts the temperature dependence of the electron density 
$n_p(T_p)$.

\subsection{	The Evolution of Stellar Flares	}

In the $EM_p-T_p$ scaling law (Eq.~11), the
observables represent the peak emission measure $EM_p=EM(t=t_p,T=T_p)$ and peak
temperature $T_p=T(t=t_p)$ measured at the peak time $t=t_p$ of stellar flares.
However, since both observables $EM(t)$ and $T(t)$ change during the flare as a function
of time, we inquire how closely the temperature $T_S$ predicted by
the statistical scaling law (Eq.~11),
\begin{equation}
        T_S = T_0 \left({EM_p \over EM_0}\right)^{0.22} \ ,
        \qquad T_0=10\ {\rm MK}, \qquad EM_0=10^{50.8}\ {\rm cm}^{-3} \ ,
\end{equation}
matches the observed peak temperature $T_p$ at the peak time $t=t_p$, and how the peak 
temperature $T_p$ and the statistically predicted peak temperature $T_S$ matches the maximum
temperature $T_M$ of observed stellar flares. For this purpose we collected
evolutionary phase diagrams of the temperature $T(t)$ versus the emission measure $EM(t)$
observed in 8 different flares (Fig.~2). The curves of 5 stellar flares observed with
{\sl GINGA} are from the stars HR1099 [or V711 Tau] (Stern 1996), 1EQ1839.6+8002 (Pan et al.~1997),
II Peg (Doyle et al.~1991), UX Ari (Tsuru et al.~1989), and Algol (Stern et al.~1992);
two observations of AB Dor flares are observed with
{\sl BeppoSAX} (Maggio et al.~2000), and one observation of Proxima Centauri was obtained
with {\sl XMM-Newton} (Reale et al.~2003). The evolutionary curves are shown
in Fig.~2, including error bars in temperature, and some are shown with 90\% confidence regions
for both derived (EM and T) parameters. The peak emission measures $EM_p$ are indicated
with a vertical dashed line in Fig.~2, ranging from $EM_p=10^{51.3}$ cm$^{-3}$ (for Proxima
Centauri) up to $EM_p=10^{55.1}$ cm$^{-3}$ (for a large flare on UX Ari). The rise time
has not been captured for two observations (II Peg and UX Ari). The observed peak
temperatures $T_p$ are also indicated, ranging from $T_p=10^{7.10}=12.5$ MK
(for Proxima Centauri) to $T_p=10^{8.06}=115$ MK (AB Dor, 1997-Nov-29). At the same
time we plot also the statistically predicted scaling law temperatures $T_S$ (using
Eq.~15) and the flare maximum temperatures $T_M$, which is usually reached shortly
before the emission measure peak. Note that the flare rise covers the upper (high-temperature)
part of the evolutionary curve, the flare peak time $t=t_p$ is defined at the right-most
datapoint (highest emission measure), and the flare decay or cooling phase extends over
the lower part of the evolutionary curve, from right to left, so the time $t$ can be
tracked in clock-wise direction along the evolutionary curve. We list the three values
of temperature $T_p$, $T_S$, and $T_M$ in Table 1 and see that there is a close match
between them, differing not more than $|log(T_S/T_p)| \lapprox 0.13$, corresponding
to a factor of $\approx 1.3$. This means that the effective flare peak temperature $T_p$ 
is predicted with an uncertainty of $\approx 30\%$ using the observed flare peak 
emission measure $EM_p$ and the statistical $EM_p-T_p$ scaling law (Eq.~11).

\section{	SOLAR FLARE OBSERVATIONS 			}

\subsection{	Flare Peak Emission Measure versus Temperature	}

In Fig.~3 we show a compilation of flare peak emission measures $EM_p$ versus
flare peak temperatures $T_p$ observed in solar flares. We included data sets
from small-scale flares (also called nanoflares) in EUV to large-scale flares
observed in soft X-rays. Small-scale flares (also called {\sl heating events 
in quiet corona}) were measured in EUV with typical temperatures of 
$T\approx 1.0-1.5$ MK (derived from 171/195 \ang\ filter ratios)
and total emission measures of $EM_p \approx 10^{43}-10^{45}$ cm$^{-3}$ 
(Krucker \& Benz 2000; Aschwanden et al.~2000a). 
The next larger category of small flares is observed in soft X-rays, also called
{\sl active region transient brightenings}, for which Shimizu (1995) measured
typical peak emission measures of $EM_p \approx 10^{44.5}-10^{47.5}$ cm$^{-3}$
and peak temperatures of $T_p \approx 4-8$ MK with {\sl Yohkoh/SXT}.
For large solar flares, peak emission measures in the range of
$EM_p \approx 10^{45}-10^{50}$ cm$^{-3}$ and flare peak temperatures in the
range of $T_p\approx 6-30$ MK were measured, using observations from
{\sl Skylab} (Pallavicini, Serio, \& Vaiana 1977), 
{\sl GOES} (Feldman et al.~1995a, 1996; Garcia 1998),
{\sl Yohkoh/BCS} (Sterling et al.~1997), and
{\sl RHESSI} (Battaglia, Grigis, \& Benz 2006).  
Comparison of flare temperatures simultaneously measured with {\sl GOES}
and {\sl RHESSI} reveal a systematic bias that RHESSI determines higher
temperatures, i.e., $T_{RHESSI} \approx 1.31\ T_{GOES} + 3.12$ [MK], which 
indicates that {\sl RHESSI} fits only the high-temperature tail 
(at $\ge 3$ keV), while the {\sl GOES} temperatures are weighted by the peak
of the emission measure distribution (Battaglia et al.~2006). Combining all
these measurements together (except for RHESSI, which has a particular 
high-temperature bias) and performing a linear regression fit in the entire 
temperature range of $T_p\approx 1-30$ MK with the {\sl ordinary least square 
bisector method} we find a statistical correlation (Fig.~3) of
\begin{equation}
        EM_p(T_p) = 10^{48.4} \left({T_p \over 10\ {\rm MK}}\right)^\alpha  
			\ [ {\rm cm}^{-3}] \ , \quad \alpha = 4.7 \pm 0.1 \ .
\end{equation}
Note that the powerlaw slope of the $EM_p-T_p$ relation is essentially identical
for both the solar and stellar flare data sets, but stellar flares have a higher
temperature range ($T_p \approx 10-150$ MK) and have about a factor of 
$10^{50.8-48.4} \approx 250$ (comparing the factors in Eqs.~11 and 16) 
higher emission measures in the overlapping temperature range ($T_p\approx 10-30$ MK). 

\subsection{	Flare Duration versus Temperature 		}

In Fig.~4 we show a compilation of flare durations $\tau_f$ versus
flare peak temperatures $T_p$ observed in solar flares. The flare durations
were obtained from $\tau_f = (t_{end}-t_{start})$ of 12 TRACE nanoflares 
(Fig.~9 of Aschwanden et al.~2000a), 
from $\tau_f=\tau_{total}/n_{peaks}$ of 23 {\sl SoHO/EIT} quiet-Sun brightening 
events (Table 1 of Krucker \& Benz 2000), 
from $\approx 200$ active region transient brightening events
observed with {\sl Yohkoh/SXT} (Fig.~3 in Shimizu 1995),
from the total durations $\tau_T$ of 31 flares observed with {\sl Skylab}
(Table 3 of Pallavicini et al.~1977), 
from $\tau_f=(\tau_{rise}+\tau_{decay})$ of 14 flares observed with
{\sl Yohkoh/SXT} (Table 6 of Garcia 1998),
from $\tau_f=(t_{end}-t_{start})$ of 9 GOES light curves (Sterling et al.~1997), 
and from 19 flares observed with {\sl Yohkoh/SXT} (Table 1 of Metcalf \& Fisher 1996). 
The EUV nanoflares have a typical duration of 
$\tau_f \approx 5\times 10^2 - 2 \times 10^3$ s ($\approx
7-33$ min), while the larger flares have typical durations of 
$\tau_f \approx 5\times 10^2-3\times 10^4$ s ($\approx$10 min - 10 hrs). Solar
and stellar flares have comparable durations in the overlapping temperature
range of $T\approx 10-30$ MK. 
Fitting all solar and stellar flare durations
combined with the {\sl ordinary least square bisector method} we obtain a statistical
correlation (Fig.~4) of 
\begin{equation}
        \tau_f(T_p) = 10^{3.4} \left({T_p \over 10\ {\rm MK}}\right)^\beta  
			\ [s] \ , \quad \beta = 0.91 \pm 0.05 \ .
\end{equation}
which encompasses 67\% of the flare durations within a factor of $\approx 3$.

\subsection{	Flare Length Scales versus Temperature 			}

We can measure geometric parameters of flares only in solar data where we have
spatial resolution using EUV and soft X-ray imagers. The most directly measured
geometric parameter is the length $L$ or area $A$ of a flare, while the flare 
volume $V$ can only be inferred indirectly from the flare area.
We compile measurements of the flare length scales $L$ versus
flare peak temperature $T_p$ in Fig.~5, where we quote either the observerd
(projected) loop lengths $L$, or length scales converted from the measured area $A$,
i.e., $L = A^{1/2}$, or quoted volume, i.e., $L \propto V^{1/3}$.

Flare length scales were measured in the range of:
$L=2-11$ Mm for 12 nanoflares observed with {\sl TRACE} ($L=\sqrt{l\times w}$ in
Table 1 of Aschwanden et al.~2000a), 
$L=4-14$ Mm for 23 quiet-Sun brightenings
observed with {\sl SoHO/EIT} (from $L=\sqrt{A}$ in Table 1 of Krucker \& Benz 2000),
$L=3-36$ Mm for 23 soft X-ray bright points
observed with {\sl MSSTA} ({\sl Multispectral Solar Telescope Array}) (half
lengths $L$ in Table 1 of Kankelborg et al.~1997), 
$L\approx 5-40$ Mm in $\approx 200$ transient soft X-ray brightenings observed with
{\sl Yohkoh/SXT} (Fig.~4 in Shimizu 1995), 
$L=5-54$ Mm in 31 flares observed with {\sl Skylab} (from $L=V^{1/3}$ 
in Table 3 of Pallavicini et al.~1977), 
$L=12-144$ Mm for 14 flares observed with {\sl Yohkoh/SXT} 
(observed loop lengths $L_{obs}$ in Table 6 of Garcia 1998), 
$L=3-13$ Mm in 20 soft X-ray flares observed with {\sl Yohkoh/SXT} 
(Table 1 in Reale et al.~1997), 
and $L=3-31$ Mm in 19 soft X-ray flares observed with {\sl Yohkoh/SXT}
(Table 1 in Metcalf \& Fisher 1996). 
Fitting all solar length scales combined with the {\sl ordinary least square 
bisector method} we obtain a statistical correlation (Fig.~5) of 
\begin{equation}
        L (T_p) = 10^{9.4} \left({T_p \over 10\ {\rm MK}}\right)^\beta  
			\ [cm] \ , \quad \beta = 0.91 \pm 0.04 \ .
\end{equation}
which encompasses 67\% of the flare durations within a factor of $\approx 3$.
 
\section{	THEORETICAL MODELING OF SCALING LAWS		}

\subsection{	The Standard RTV Law 				}

A scaling law between the peak temperature $T_{max}$, pressure $p$, and loop length $L$
in static coronal loops has been derived by Rosner, Tucker, \& Vaiana (1978), assuming an
equilibrium between a spatially uniform heating rate $(E_H)$ and the conductive 
($E_{cond}$) and radiative ($E_{rad}$) loss rates, the so-called {\sl RTV} scaling law,
\begin{equation}
	T_{max}(p, L) \approx 1400 (p L)^{1/3} \ .
\end{equation}
While the validity of this scaling law applies 
to a coronal loop in hydrostatic equilibrium, it might also approximately apply to a
flaring loop near the peak time, because both (1) the energy and (2) momentum equations
are nearly balanced near the flare peak.

(1) Energy equation: In the initial heating phase, 
the heating term is larger than the combined conductive and radiative loss term,
i.e., $E_H > (E_{cond}+E_{rad})$,
because the average flare temperature is rising ($dT(t)/dt > 0$) due to excessive heating.
After the peak time, the conductive and radiative losses exceed the decreasing or
stopped heating rate, 
i.e., $E_H < (E_{cond}+E_{rad})$,
because the flare plasma is cooling ($dT(t)/dt < 0$).
Therefore, there is a balance between heating and loss terms,
i.e., $E_H \approx (E_{cond}+E_{rad})$,
near the flare peak time $t_p$ [defined by the peak in emission measure $EM_p=EM(t=t_p$)]. 
Hydrodynamic simulations by Jakimiec et al.~(1992) show that the RTV scaling law
predicts a maximum electron density $n_p$ that can be reached in a flare loop
if a constant heating rate is applied sufficiently long after the maximum
temperature $T_p$ is reached. In Appendix A we show how the agreement between
the maximum electron density $n_p$ and the RTV-predicted value scales with 
the heating duration, but is nearly independent of the maximum temperature $T_p$
and heating rate $E_H$. 

(2) Momentum equation: Secondly, also the momentum equation is nearly balanced after 
a flare loop is filled. The losses are dominated by thermal
conduction at high flare temperatures (say at $T\gapprox 10$ MK; Aschwanden
\& Alexander 2001), and the loop filling time is much shorter (in the order of $\tau_{fill}
\lapprox 1$ min; see hydrodynamic simulations by
MacNeice et al.~1984; Nagai \& Emslie 1984; Fisher et al.~1985a,b,c;
Mariska \& Poland 1985; Yokoyama \& Shibata 1998, 2001; Hori et al.~1997, 1998)
than the conductive or radiative cooling time (in the order of $\tau_{cool}
\gapprox 10$ min; Antiochos \& Sturrock 1978; Culhane et al.~1994; Aschwanden
\& Alexander 2001), and thus pressure gradients resulting from the chromospheric
evaporation process are largely balanced out so that the momentum equation is
approximately fulfilled. Also the assumption
of a constant pressure (made in the RTV law) is better fulfilled in the hot
soft X-ray emitting flare loops (because of the larger pressure scale heights, 
$\lambda_T > L$) than in the cooler EUV-emitting coronal loops (where often 
$\lambda_T < L$). 

Applying the standard RTV scaling law to flare loops now, 
with $T_p \approx T_{max}$, by inserting the expression for the thermal pressure,
\begin{equation}
	p(n_p, T_p)	= 2 n_p k_B T_p \ ,
\end{equation}
we obtain a scaling law for the peak density $n_p$ as a function of the 
peak temperature $T_p$ and loop length $L$,
\begin{equation}
        n_p(T_p, L) = c_0 \ {T_p^2 \over L} \ , \quad
		      c_0 = {1 \over 1400^3 \ 2 k_B} \approx 1.3 \times 10^6 
			  \ [{\rm cm}^{-2}  {\rm K}^{-2}] \ .
\end{equation}
where we defined a constant $c_0$ for the numerical factor.
The electron density $n_p$, however, is not a directly measured quantity
in most solar and stellar flare observations, so instead we use the directly 
measured quantity of the total (volume-integrated) emission meausure $EM_p$,
\begin{equation}
	EM_p(n_p, V) = n_p^2 V \ ,
\end{equation}
involving the total volume $V$ of all flare loops. The standard RTV law, expressed 
in terms of the total emission measure $EM_p$, is then
\begin{equation}
        EM_p(T_p, L, V) = n_p^2(T_p,L)\ V = c_0^2 \ {T_p^4 \ V \over L^2} \ .
\end{equation}								%Eq.(23)

\subsection{	Serio's Scaling Law	}

It is known that the RTV law underestimates the densities of active region loops
observed in EUV (at temperatures of $T\approx 1-2$ MK), as a result of the
(sometimes invalid) assumptions of a spatially uniform heating rate and constant 
pressure (Lenz et al.~1999; Aschwanden, Nightingale, \& Alexander 2000b; 
Aschwanden, Schrijver, \& Alexander 2001; Aschwanden et al.~2003;
Winebarger et al. 2003; Aschwanden et al.~2007). Applying Serio's
scaling law (1981), which generalizes the RTV law by including a non-uniform
heating scale height $s_H$ and gravitional stratification, leads to the following
correction factor $q_{Serio}$, 
\begin{equation}
	T_{max}(p, L) = 1400 (p L)^{1/3} \ q_{Serio} \ ,		%Eq.(24)
\end{equation}
\begin{equation}
	 q_{Serio}(L, T_p, s_H) = 
	\exp{\left( -0.08 {L\over s_H} - 0.04 {L\over \lambda_p}\right)} \ .
\end{equation}								%Eq.(25) 
where $L$ is the loop half length, $s_H$ the heating scale height, and
$\lambda_p$ the pressure scale height. 
This correction factor depends on $(L, T_p, s_H)$, since the pressure scale height
is a function of the temperature, $\lambda_p \approx 47 \times T_{MK}$ [Mm]. In order
to apply Serio's scaling law we can simply replace the constant $c_0$ with a new
function $c_1(L, T_p, s_H)$,
\begin{equation}
	c_1(L, T_p, s_H) = {c_0 \over q_{Serio}^3}
		= c_0 \exp{\left( 0.24 {L\over s_H} + 0.12 {L\over \lambda_p}\right)} \ .
\end{equation}								%Eq.(26)
If we insert Serio's correction factor $c_1=c_0/q_{Serio}^3$ (Eq.~26) into
the RTV law of the total emission measure (Eq.~23) we obtain the following scaling,
\begin{equation}
        EM_p(T_p, L, V, s_H) = c_1^2 \ {T_p^4 \ V \over L^2} 
		= c_0^2 {T_p^4 \ V \over L^2} 
		\exp{(0.48 {L\over s_H} + 0.24 {L \over \lambda_p})} \ .
\end{equation}								%Eq.(27) 
Stellar flares have temperatures of $T_p \gapprox 10$ Mm, and thus pressure scale 
heights of $\lambda_p = 47$ Mm $\times (T_p$/1 MK) $>$470 Mm that are much larger
than the expected solar loop lengths (typically $L_0=25$ Mm at $T_0$=10 MK). We 
can therefore assume $L \ll \lambda_p$ and neglect the second term $\exp(L/\lambda_p)$ 
in Serio's correction and can approximate (Eq.~27) by
\begin{equation}
        EM_p(T_p, L, V, s_H) \approx c_0^2 {T_p^4 \ V \over L^2} \exp{(0.5 {L\over s_H})} \ .
\end{equation}
which yields higher emission measures than the RTV law (Eq.~23) 
for heating scale heights of $s_H \lapprox L$. However, for relatively short heating
scale heights, say $s_H/L \lapprox 1/3$, a density inversion occurs at the temperature
maximum of the loop, which is unstable against the Rayleigh-Taylor instablity
($\nabla n \cdot {\bf g} < 0$). This is confirmed by analytical calculations as well as
with hydrodynamic simulations, where a lower limit is found for stable hydrostatic equilibria
at $s_H/L \lapprox 1/3$ (Serio et al.1981; Aschwanden \& Schrijver 2002;
Winebarger et al.~2003). Given this instabilitiy limit we find that Serio's correction
factor can increase the emission measure $EM_p$ only by a factor of 
$\exp(0.5 \times 3) \approx 4$, or the electron density $n_e \propto \sqrt{EM}$ by a
factor of $\approx 2$, which is also confirmed by hydrodynamic simulations (Winebarger et al.~2003).  

\subsection{	Fractal Scaling of Flare Volume  			}

In early models, stellar flares were modeled with a single loop, so that
the volume $V=L\times a$ was described by a loop length $L$ and a cross-sectional area $a$.
If this cross-sectional area $a$ is chosen as a constant for small and large flares,
the flare volume would just scale linearly with the length scale $L$, i.e., $V \propto L^1$.
However, we think that such geometries are unrealistic, given the fact that large solar flares
always reveal arcades with hundreds of flare loops, remnants of multiple magnetic 
reconnection sites. A uniformly filled flare arcade is expected to scale with $V\propto L^3$
in the Euclidian limit. 
In arcsecond high-resolution images such as from {\sl TRACE}, however, it becomes 
evident that the flare region is not uniformly filled, but rather has a filamentary 
structure that can be described with a fractal dimension or an area filling factor.
A detailed study (Aschwanden \& Aschwanden 2006a) 
of 20 GOES X- and M-class flares observed 
with TRACE has revealed that the fractal area (normalized by
the time-integrated flare area $A_f$) varies from near zero at the beginning
of the flare to a maximum of $A(t)/A_f = 0.65 \pm 0.12$ at the peak time of the flare,
which corresponds to an {\sl area fractal dimension} of $D_A \lapprox 1.89\pm0.05$ 
(at the flare eak time), also called {\sl Haussdorf dimension} $D_A$, which defines 
the scaling of a fractal area $A$ with length scale $L$, i.e., $A(L)\propto L^{D_A}$.
Also a statistical study of nanoflares has shown a Haussdorf dimension of 
$D_A = 1.5\pm0.2$ (Aschwanden \& Parnell 2002). 
From the measured {\sl area fractal dimensions} $D_A$ at the flare peak time,  
a {\sl volume fractal dimension} $D_V \approx 2.37 \pm 0.14$ was derived for 
a flare model with an arcade geometry (Aschwanden \& Aschwanden 2006b),
where the fractal flare volume scaling is defined by $V(L) \propto L^{D_V}$. 
The resulting volume filling factors were found to be in the range of
$q_V \approx 0.03-0.08$ at the flare peak time. 

The implication of the previous studies is that the flare volume has a fractal
scaling of $V(L) \propto L^{2.4}$, rather than Euclidian filling with $V(L) \propto L^3$,
which affects our scaling laws. We can quantify this fractal volume scaling by
\begin{equation}
	V(L) = q_V L_0^3 \left({L \over L_0}\right)^{D_V} \ ,		
\end{equation}								 
where $L_0=10^{9.4}$ cm (=25,000 km) is the average length scale of a solar flare
loop (at a typical flare temperature of $T_p=10$ MK, according to Eq.~18),
$q_V=0.03-0.08$ is the average volume filling factor, and $D_V \approx 2.37\pm0.14$
is the volume fractal dimension. This definition fulfills the normalization 
$V(L=L_0)=q_V L_0^3$, so the average volume filling factor is defined by $q_V=V/V_0$ at
$V_0=L_0^3$. 

Inserting this volume scaling $V$ into the RTV law (Eq.~23) we obtain the following
relation,
\begin{equation}
        EM_p = c_0^2 \ {T_p^4 \ V \over L^2} = 
	       c_0^2 q_V L_0 T_p^4 \left({L \over L_0}\right)^{D_V-2}
		\propto T_p^4 L^{0.4} \ .		
\end{equation}								
Therefore, comparing total emission measures of solar and stellar flares at the same
flare peak temperature, we expect the following dependence
\begin{equation}
        {EM_p^* \over EM_p^\odot} = {q_V^* \over q_V^\odot} 
		\left({L* \over L^\odot}\right)^{D_V-2} \ .
\end{equation}								%Eq.(28)
while the relative densities scale reciprocally to the loop length $L$ (for the same
flare peak temperature) according to the RTV law (Eq.~21),
\begin{equation}
        {n_p^* \over n_p^\odot} = \left({L^* \over L^\odot}\right)^{-1} \ .
\end{equation}								%Eq.(29)
Is this fractal volume model consistent with the scaling laws we found for solar flares?
Inserting the observed scaling of $L(T_p) \approx T_p^{0.9}$
(Eq.~18 and Fig.~5) into the relation for the total emission measure (Eq.~30) 
we find the following scaling with the flare peak temperature $T_p$, 
\begin{equation}
        EM_p \propto T_p^4 \ [L(T_p)]^{D_V-2} \propto T_p^{4.3} \ .
\end{equation}								%Eq.(30)
which is close to the observed scaling of 
$EM_p(T_p) \propto T_p^{4.7\pm0.1}$ for solar flares and 
$EM_p(T_p) \propto T_p^{4.5\pm0.4}$ for stellar flares (Fig.~3), 
and thus the fractal scaling is consistent with observations.
This RTV model (Eq.~21) predicts also the following dependence of the density $n_p$ with 
flare peak temperature $T_p$, using the scaling $L(T_p) \propto T^{0.9}$ 
observed in solar flares,
\begin{equation}
	n_p(T_p) = c_0 {T_p^2 \over L(T_p)} 
	         = 10^{10.7} \left({T_p \over 10\ {\rm MK}}\right)^{1.1} 
		\ [{\rm cm}^{-3}] \ .
\end{equation}								%Eq.(31)
Thus, we predict for stellar flares in the temperature range of $T_p\approx 10-100$ MK
up to an order of magnitude higher electron densities than for solar flares in the
temperature range of $T_p\approx 10-20$ MK.

\subsection{	Flare Conductive Cooling Times	  		}

For the physical process of conductive cooling, there
is a simple relation between the maximum temperature $T_{max}$ and loop length $L$
for a coronal loop in hydrostatic equilibrium,
\begin{equation}
        {dE_{cond} \over dt\ dV}(T,L) = {d \over dT}[-\kappa T^{5/2} {dT\over ds}]
                = -{2 \over 7} \kappa {dT^{7/2} \over ds} \approx 
		  -{2 \over 7} \kappa {T^{7/2} \over L^2} \ ,
\end{equation}
where $\kappa = 9.2 \times 10^{-7}$ [erg s$^{-1}$ cm$^{-1}$ K$^{-7/2}$] 
is the classical Spitzer conductivity. We can argue that the heating
rate is dominated by conductive losses at the flare peak time,
if the flare is sufficiently hot so that radiative losses can be neglected. For solar flares
it has indeed been shown that conductive cooling always dominates over radiative losses
in the hot phase of flares, say at temperatures of $T \gapprox 10$ MK (e.g.,
Antiochos \& Sturrock 1978; Culhane et al.~1994; Aschwanden \& Alexander 2001).
Assuming this balance of heating and conductive loss at the flare peak time,
i.e., $dE_H/dt\ dV \approx dE_{cond}/dt\ dV$, leads (by inserting the first RTV law
of Eq.~19 into Eq.~35) directly to the second RTV scaling law for the heating rate 
(Rosner et al.~1978),
\begin{equation}
        {dE_H \over dt\ dV} \propto -{dE_{cond} \over dt\ dV} \propto 
	{T^{7/2} \over L^2} \propto {p^{7/6} \over L^{5/6}} \ . 
\end{equation}
The thermal energy of the flare plasma per volume is, 
\begin{equation}
        {dE^T \over dV}  = 3 n_p k_B T_p \ ,
\end{equation}
which can be expressed in terms of flare peak temperature $T_p$ and flare length
scale $L$ (by substituting the first RTV law Eq.~21), 
\begin{equation}
        {dE^T \over dV}  = 3 k_b c_0 \ \left( {T_p^3 \over L} \right) \ .
\end{equation}
The two expressions of the thermal energy (Eq.~38) and the conductive cooling rate 
(Eq.~35) allow us to define the cooling time $\tau_c$ by thermal conduction,
\begin{equation}
        \tau_c(T_p, L) = {dE^T/dV \over dE_{cond}/dt\ dV} =  
	{ 3 k_b c_0 \over 2 \kappa /7} {(T_p^3 /L) \over (T_p^{7/2}/L^2)} 
	\approx 2 \times 10^{-3} \ {L \over T_p^{1/2}} \ .
\end{equation}
If we express this relation in dimensionless units in terms of the reference values
$T_0=10^7$ K and $L_0=25$ Mm we obtain the following scaling law, 
\begin{equation}
        \tau_c(T_p, L) = \tau_{c0} \ \left({ L \over L_0 }\right)
				  \left({ T_p \over T_0 }\right)^{-1/2} \approx
                           1600 \ \left({ L \over 25\ {\rm Mm}}\right)
				  \left({ T_p \over 10 \ {\rm MK} }\right)^{-1/2} 
				   \ [s] \ .
\end{equation}
In Fig.~6 (top left panel)
we plot this theoretically estimated cooling time $\tau_c(T_p, L)$
calculated from the observed values of the flare peak temperature $T_p$ and
flare length scale $L$ as a function of the observed flare duration times
$\tau_f$. We find that the theoretically estimated conductive cooling time
$\tau_c$ of most flares is comparable or shorter than the flare
duration $\tau_f$.

\subsection{	Flare Radiative Cooling Times	  		}

The radiative loss rate is generally expressed as a product of the electron
density $n_e$, ion density $n_i$, and the radiative loss function $\Lambda (T)$,
which in the coronal approximation (with full ionization, i.e., $n_e \approx n_i$) is
\begin{equation}
	{dE_{rad} \over dV \ dt}(n_e, T) = n_e^2 \Lambda(T) \ ,
\end{equation}
where the radiative loss function can be approximated by piece-wise powerlaws
(Rosner et al.~1978; Mewe et al.~1985),
\begin{equation}
    \Lambda(T) = \left\{
	\begin{array}{ll}
	10^{-21.94}	     & {\rm for}\ 10^{5.75} < T < 10^{6.3} \ {\rm K} \nl
	10^{-17.73} T^{-2/3} & {\rm for}\ 10^{6.3} < T < 10^{7.3}  \ {\rm K} \nl
	10^{-24.66} T^{1/4}  & {\rm for}\ T > 10^{7.3} \ {\rm K} 
	\end{array}
    \right. \ .
\end{equation}
The radiative cooling time can then be defined as the ratio of the thermal energy
(Eq.~39) and the radiative loss rate (Eq.~41), where we can eliminate the unknown
density by inserting the RTV law (Eq.~21),
\begin{equation}
        \tau_r(T_p, L) = {dE^T/dV \over dE_{rad}/dt\ dV} =  
	{ 3 k_b c_0 \over c_0^2 \Lambda(T_p)} {(T_p^3 /L) \over (T_p^4/L^2)} = 
	{ 3 k_b \over c_0 \Lambda(T_p)} {L \over T_p} \ , 
\end{equation}
which reads in dimensionless units,
\begin{equation}
        \tau_r(T_p, L) = \left\{ 
	\begin{array}{ll}
		 700 \ \left({ L \over 25\ {\rm Mm}}\right)
			\left({ T_p \over 10 \ {\rm MK} }\right)^{-1} 
			& {\rm for}\ 10^{5.75} < T < 10^{6.3} \ {\rm K} \nl
		2000 \ \left({ L \over 25\ {\rm Mm}}\right)
			\left({ T_p \over 10 \ {\rm MK} }\right)^{-1/3} 
			& {\rm for}\ 10^{6.3} < T < 10^{7}    \ {\rm K} \nl
		6500 \ \left({ L \over 25\ {\rm Mm}}\right)
			\left({ T_p \over 10 \ {\rm MK} }\right)^{1/4} 
			& {\rm for}\ T > 10^{7.3}     \ {\rm K}
		\end{array}
	\right.  \ [s] \ .
\end{equation}
We plot the radiative cooling times $\tau_r(T_p, L)$ calculated with the observed
values $(T_p, L)$ in Fig.~6 (top right panel)
as a function of the flare duration $\tau_f$. These
radiative cooling times based on the RTV scaling law are generally longer
than the conductive cooling times, and they clearly exceed the flare durations for
most of the flares, up to an order of magnitude for the short EUV flares with
durations of $\tau_f \lapprox 10^3$ s. Since the observed flare duration should be
an upper limit of the cooling time, either the conductive or radiative cooling time
should be equal or shorter. A combined cooling time $\tau$ can be defined from the
exponential folding time that would result from the product of the two 
exponential cooling processes,
\begin{equation}
	{1 \over \tau} = {1 \over \tau_c} + {1 \over \tau_r} \ .
\end{equation}
We plot this combined cooling time $\tau$ as a function of the flare duration
in Fig.~6 (bottom left panel) using the RTV law. The so-defined combined 
cooling time is almost always shorter than the observed flare duration. 

Including Serio's correction factor (Eq.~25) for the conductive and radiative 
cooling times,
\begin{equation}
        \tau_c^{Serio} = \tau_c^{RTV} {c_1 \over c_0}
                        = \tau_c^{RTV} {1 \over q_{Serio}^3} \ ,
\end{equation}
\begin{equation}
        \tau_r^{Serio} = \tau_r^{RTV} {c_0 \over c_1}
                        = \tau_r^{RTV} q_{Serio}^3 \ .
\end{equation}
yields only small corrections, shown for the five solar datasets in Fig.~6.
The average correction values are $<q_{Serio}>=0.90-0.98$ for the EUV
datasets (Aschwanden et al.~2000a; Krucker \& Benz 2000), and
$<q_{Serio}>=0.05-0.95$ for the soft X-ray datasets (Pallavicini et al.~1977;
Garcia 1998; Metcalf \& Fisher 1996).
We plot the corrected flare cooling times predicted by Serio's scaling law
in Fig.~6 (bottom right panel). The major effect of Serio's correction is
that the cooling times of the larger and hotter soft X-ray emitting flare loops
become shorter, limiting essentially all flare loop cooling times to
$\tau \lapprox 10^3$ s. Thus, large flares that last significantly longer
(up to $\tau_f \lapprox 2 \times 10^4$ s) must consist of multiple subflares.

\section{	DISCUSSION  				}

\subsection{	Scaling Laws of Solar versus Stellar Flares	}

The combined $EM-T$ diagram for solar and stellar flares (Fig.~3) shows some
similarities but also intriguing differences in the scaling behavior. 
Solar flares have been observed mostly in the temperature range of
$T_p \approx 7-30$ MK (or $T_p \approx 1-30$ MK if we include the EUV 
nanoflares), while stellar flares have been detected within the
temperature range of $T_p \approx 10-150$ MK. The lack of observations
of cooler stellar flares $T_p \lapprox 10$ MK is likely to be due to the
sensitivity limit, which is about at $EM_p \gapprox 10^{51}$ cm$^{-3}$
for stellar flares. The sensitivity limit also systematically increases
with higher flare temperatures, up to $\approx 10^{53}$ cm$^{-3}$ for
the hottest stellar flares with $T_p \gapprox 100$ MK, which is likely a 
consequence of the decreasing sensitivity of current soft X-ray detectors 
at higher temperature lines. 

What is similar for both solar and stellar flares is the overall slope
of the $EM_p-T_p$ relationship, which was found to have a powerlaw
slope of $\alpha=4.7\pm0.1$ for solar flares (excluding the RHESSI
data that have a high-temperature bias), and a slope of 
slope of $\alpha=4.5\pm0.4$ for stellar flares (Fig.~3). Both powerlaw
slopes are consistent with the theoretically expected scaling of 
$\alpha \approx 4.3$ (Eq.~33), which is based on the RTV scaling law (Eq.~23),
the fractal volume scaling (Eq.~29),
and the observed spatial scaling $L(T_p)\approx T_p^{0.9}$ in solar flares
(Eq.~18). This agreement supports the assumption of the fractal volume
scaling. If we were to assume a monolithic single-loop model 
with constant cross-section that scales with $V(L) \propto L$, the
resulting scaling law would be $EM_p \propto T_p^{3.9}$, or a monolithic
cubic model with $V(L) \propto L^3$ would yield $EM_p \propto T_p^{5.9}$,
which are both less consistent with the observations. 

What is different between solar and stellar flares is the emission 
measure in the same temperature range. In the overlapping temperature
range of $T_p \approx 10-30$ MK we find that the total emission measure
of stellar flares is larger by an average factor of 
\begin{equation}
	{EM_p^* \over EM_p^\odot} = {10^{50.8} \over 10^{48.4}} \approx 250 \ ,
\end{equation}
comparing the factors (Eqs.~15, 16) of the two linear regression fits in Fig.~3. 
If we compare the theoretical model of the RTV law and the fractal volume
scaling (Eq.~31) for the same temperature range (Eq.~33), we see that the 
total volume-integrated emission measure ratio of stellar to solar flare 
depends on the volume filling factor $q_V$, length scale $L$, and volume
fractal dimension $D_V$. Since the sensitivity limit of stellar soft X-ray 
detectors represents a bias for larger emission measures, 
we think that the observed stellar flares have a bias
for both higher volume filling factors $q_V$ and flare sizes $L$. For solar
flares we found spatial filling factors in the range of $q_V=0.03-0.08$
(Aschwanden \& Aschwanden 2006b), so stellar filling factors can be up to
a factor of $q_V^*/q_V^\odot \lapprox 10-30$ higher, ameliorating the size 
requirement to be $(L^*/L^\odot)^{D_V-2} \gapprox 10-25$ to match the 250
(Eq.~48) times larger emission measures of of detected stellar flares . 

\subsection{	The Electron Density in Solar and Stellar Flares 	}

The electron density $n_p$ of the flare plasma cannot directly be measured and
is therefore dependent on the volume model if derived from the total
emission measure. If the density is naively derived from an Euclidian flare
volume, i.e., $n_p = \sqrt{EM_p / V}$, we obtain only a lower limit. 
However, detailed measurements of the area fractal dimension and modeling of the volume
fractal dimension has yielded volume filling factors of $q_V\approx 0.03-0.08$
for solar flares, which raises the average electron densities in the flare loops
(at the flare peak time) by a factor of $1/\sqrt{q_V} \approx 4-6$. Thus, a more
realistic estimate of the plasma density in flare loops, based on the observed
total emission measure $EM_p$ and flare length scale $L$ is (using the fractal
scaling of Eq.~29)
\begin{equation}
	n_p^{obs}(EM_p, L) = \left({EM_p \over q_V L_0^3 (L/L_0)^{D_V}}\right)^{1/2}	\ ,
\end{equation}
where the volume filling factor is typically in the range of $q_V\approx 0.03-0.08$
and the fractal dimension is $D_V \approx 2.4$
according to measurements of fractal flare areas (Aschwanden \& Aschwanden 2006a,b).  

On the other hand we can estimate the plasma density theoretically, using
the RTV law (Eq.~21) applied to the observed flare peak temperature $T_p$ and
loop half length $L$, 
\begin{equation}
	n_p^{RTV}(T_p, L) = c_0 {T_p^2 \over L}  \ .
\end{equation}
We show the two estimated electron densities versus each other in Fig.~7
and find a good agreement within less than an order of magnitude. Both
methods yield flare densities in the range of $n_e\approx 10^9-10^{12}$
cm$^{-3}$ for solar flares. 

Plugging in the observed scaling law for loop lengths versus flare peak
temperature, i.e., $L(T_p) \propto T_p^{0.9}$ (Eq.~18), we find an approximate 
prediction for the electron density as a function of the flare peak
temperature alone, i.e., $n_e(T_p) \propto T_p^{1.1}$ (Eq.~34), as it
can be seen in Fig.~8 by comparing with the RTV predictions for individual flare loop
lengths. This scaling law predicts a electron 
densities of $n_e\approx 10^9-10^{10}$ cm$^{-3}$ for nanoflares at $T_p=1$ MK,
densities of $n_e\approx 10^{10}-10^{11}$ cm$^{-3}$ for typical solar flares at $T_p=10$ MK,
and densities of $n_e\approx 10^{11}-10^{12}$ cm$^{-3}$ for large stellar
flares at $T_p=100$ MK. 

Is there evidence for such higher electron densities in stellar flare loops?
Recent reviews (e.g., \S 10 in G\"udel 2004; Ness et al.~2004) 
of electron density measurements in stellar flare 
loops based on density-sensitive iron line pairs quote a typical range
of $n_e \approx 2 \times 10^{11}-2 \times 10^{13}$ cm$^{-3}$. A density of
$n_e \approx 3 \times 10^{11}$ cm$^{-3}$ and a spatial scale of 0.1 stellar
radius ($L\approx 200$ Mm) was measured in a spatially resolved flare on the
eclipsing binary Algol B (Schmitt et al.~2003). This range of observed densities
($n_e^{obs} \approx 2 \times 10^{11}-2 \times 10^{13}$ cm$^{-3}$)
brackets our theoretically predicted range. We have also to keep in mind that 
previously measured densities from stellar spectroscopy (from He-like triplets 
of OV II, Ne IX, Mg XI, Si XIII, and density-sensitive line ratios) have severe 
sensitivity limitations for densities above $\gapprox 10^{12}$ cm$^{-3}$. 
So our densities predicted by the RTV law are in the same ballpark as the
observed stellar flare densities. The RTV law seems to be a good prediction
tool, and the effects of short heating times, which can be a factor of $\approx 2$
lower than the RTV predicted densities (Appendix A; Fig.~9), largely cancel out
the effects of short heating scale heigths, which can reach up to a factor of
$\approx 2$ higher densities than predicted by the RTV law.

\subsection{	An Universal Scaling Law 			}

A correlation between the volume emission measure $EM_p=n_p^2 V$ and
the flare peak temperature $T_p$ was extended from solar flares to
stellar flares (e.g., Feldman et al.~1995b; Stern 1992; Shibata \&
Yokoyama 1999, 2002). A theoretical attempt was made to explain the
solar/stellar $EM_p-T_p$ correlation with a universal flare model
in terms of magnetic reconnection by Shibata \& Yokoyama (1999, 2002). 
Using the result of numerical MHD simulations of flares conducted
in Yokoyama \& Shibata (1998), where the flare peak temperature
scales as $T_p \propto B^{6/7} n_0^{-1/7} L^{2/7}$ (with $B$ 
the magnetic field strength and $n_0$ the electron density outside the
reconnection region), setting the thermal pressure equal to the
magnetic pressure, $2 n_p k_B T_p \approx B^2/8 \pi$, and assuming Euclidian
volume scaling, $EM_p \propto n_p^2 L^3$, they arrived at an
``universal scaling law'' of (their Eq.~5)
\begin{equation}
	EM_p \approx 10^{48} \left({B \over 50\ {\rm G}}\right)^{-5}
		\left({n_0 \over 10^9\ {\rm cm}^{-3}}\right)^{3/2}
		\left({T \over 10\ {\rm MK}}\right)^{17/2} \ [{\rm cm}^{-3}] \ ,
\end{equation}
so the emission measure scales with a power of $EM_p \propto T_p^{8.5}$.
If we introduce the fractal scaling of the flare volume, $EM_p \propto
n_p^2 L^{D_V}$, the universal scaling law of Shibata and Yokoyama (1999)
takes the following form,
\begin{equation}
	EM_p \propto B^{(4-3D_V)} n_0^{D_V/2} T_p^{(7/2)D_V-2} \ ,
\end{equation}
which yields the following coefficients for fractal scaling ($D_V=2.0$ and 
$D_V=2.4$) and Euclidian scaling ($D_V=3$),
\begin{equation}
    EM_p \propto = \left\{
	\begin{array}{ll}
	B^{-2}   n_0^{1.0} T_p^{5.0} & {\rm for}\ D_V=2.0 \nl
	B^{-3.2} n_0^{1.2} T_p^{6.4} & {\rm for}\ D_V=2.4 \nl
	B^{-5}   n_0^{1.5} T_p^{8.5} & {\rm for}\ D_V=3.0
	\end{array}
    \right. \ .
\end{equation}
Thus, if the magnetic field $B$ and electron density $n_0$ outside the
reconnection region do not have a systematic scaling with the flare peak
temperature $T_p$, the universal scaling law of Shibata and Yokoyama (1999) 
predicts a scaling law of $EM_p \propto T_p^{6.4}$ for the observed volume fractal
scaling of $D_V=2.4$, which is somewhat steeper than our measured values
of $EM_p \propto T_p^{4.5\pm0.4}$ for stellar flares, or predicted by
the RTV law, i.e., $EM_p \propto T_p^{4.3}$. Testing the validity of
the universal scaling law of Shibata and Yokoyama (1999) requires also
statistics on magnetic field strengths $B$ and ambient electron densities
$n_0$ in flares. 

\section{	CONCLUSIONS 				}

We compiled directly observed parameters from solar and stellar flares,
such as the volume peak emission measure $EM_p$, flare peak temperature $T_p$,
flare duration $\tau_f$, and flare length scale $L$ (the latter only for
solar flares). A prominent statistical correlation is found between the
volume emission measure $EM_p$ and flare peak temperature $T_p$, which
scales as $EM_p \approx T_p^{4.7}$ for both solar and stellar flares.
Another recent study demonstrated that the flare volume has a fractal scaling,
$V(L)\propto L^{2.4}$, rather than the generally used Euclidian scaling of 
$V(L)\propto L^3$. Applying the RTV scaling law, combined with the fractal 
volume scaling and the statistical $L-T_p$ correlation $L(T_p) \propto T_p^{0.9}$,
leads directly to a theoretically predicted scaling law of $EM_p \propto T_p^{4.3}$, 
which explains the observed correlations in both solar and stellar flares. 

A second result we find is an unexplained offset by a factor of about 250 
between solar and stellar flares at the same temperature, which is likely
due to a selection bias for stellar flare events with larger volume filling
factors and larger spatial scales. Interestingly, however, this selection
bias does not affect the overall $EM-T$ relationship and the lower threshold
has a similar functional dependence of $EM_{min} \propto T_p^{4.7}$, probably  
because the detector sensitivities are dropping off with a similar function
with higher temperatures. 

A third result is that our model of fractal flare volume scaling provides
realistic estimates of volume filling factors, and thus of flare densities.
We find that the electron densities in solar flare loops can be predicted
based on our fractal scaling in close agreement to the predictions of the
RTV law. The agreement of the predicted electron densities with both methods
agrees always better than an order of magnitude (Fig.~7), although the 
absolute magnitude varies by three orders of magnitude between the smallest
nanoflares and the largest solar flares, i.e., $n_e \approx 10^9-10^{12}$
cm$^{-3}$. Since the RTV scaling law (combined with the observed 
$L \propto T_p$ correlation) predicts about a linear relationship
between the electron densities and flare peak temperatures, i.e., 
$n_p \propto T_p$, we expect up to an order of magnitude higher electron
densities in the largest stellar flares due to the higher temperature than
in solar flares.

The determination of correct scaling laws allows us also to infer realistic 
estimates of the (conductive and radiative) flare cooling times, which can be 
tested from the e-folding decay time of individual peaks in (solar and
stellar) flare light curves. 

The scaling laws allow us also to eliminate temperature biases in the statistics
of the total thermal energy of flares, i.e., 
$E_T \propto n_p T_p V \propto EM_p T_p/n_p$. 
Since we find that the electron density (corrected for a fractal filling factor)
scales approximately as $n_p \propto T_p^{1.1}$, the thermal energy scales
approximately as $E_T \propto EM_p$, and thus the observed total emission measure
$EM_p$ can be used as a good proxy for the thermal flare energy $E_T$.
Such unbiased frequency distributions of flare energies $N(E_T)$ 
permit us then to determine whether there is more energy in large or small flares, 
an important test for nanoflare heating theories. 

\acknowledgements
We acknowledge very helpful comments from the referee, Kazunari Shibata, Marc Audard, 
RHESSI data from Marina Battaglia, and hydrodynamic simulations from David Tsiklauri.
Part of this work was supported by NASA contract NAS5-38099 (TRACE mission) and the 
NASA/LWS-TRT contract ``Energy Scaling of Flares'' NAG5-13490.  

\appendix
\section{ DENSITY COMPARISON OF RTV-LAW WITH HYDRODYNAMIC SIMULATIONS }

Here we compare the electron densities predicted by the RTV law (Eq.~19;
\S 4.1) with maximum flare densities $n_{max}$ obtained from hydrodynamic
simulations of heated flare loops. We make use of a parametric study of
radiative hydrodynamic modeling conducted in Tsiklauri et al.~(2004),
where a flare loop with a fixed half length of $L=55$ Mm was heated
with variable heating functions, specified by 5 different heating
rates ($E_{H0}=$0.60, 3.0, 15.0, 30.0, and 60.0 erg cm$^{-3}$ s$^{-1}$)
applied at the flare loop apex for 4 different (Gaussian) durations
($t_{heat}$ = 41, 82, 164, 329 s). We show 4 evolutionary curves of
the apex temperature $T_{apex}(t)$ as a function of the electron density
$n_e(t)$ for the lowest and highest heating rate, as well as for the
shortest and longest heating duration in Fig.~9. We see that the 5
different sets of heating rates produce approximately maximum flare
temperatures of $T_{max} \approx$ 14, 22, 35, 43, and 53 MK. All
parametric values are also tabulated in Table 1 of Tsiklauri et al.~(2004).
For each of the 20 evolutionary curves we mark the maximum values 
$n_{max}$ and $T_{max}$ of the simulated flare loops with a diamond
symbol in Fig.~9 and compare it with the RTV law (Eq.~21), which is
indicated with a dashed line in Fig.~9. We can consider now the
density ratio $n_{max}/n_{RTV}$ at the flare maximum temperature and
find that this ratio is a systematic function of the heating time scale
$t_{heat}$, but is almost independent of the heating rate $E_{H0}$ and
maximum temperature $T_{max}$. This ratio amounts to 
$n_{max}/n_{RTV}=1.09\pm0.06$ for $t_{heat}=328$ s, 
$n_{max}/n_{RTV}=0.81\pm0.07$ for $t_{heat}=164$ s, 
$n_{max}/n_{RTV}=0.57\pm0.06$ for $t_{heat}=82$ s, 
$n_{max}/n_{RTV}=0.35\pm0.06$ for $t_{heat}=41$ s, 
so the RTV overpredicts the maximum density for small heating time scales,
but agrees quite well for longer heating time scales and thus  
provides a good proxi to predict the maximum flare densities.
Applying the RTV law to scaling laws of flare parameters, one has to correct
for a numerical factor for shorter heating times, but this factor is largely
independent of the flare temperature.

\clearpage

%%%%%%%%%%%%%%%%%%%%%%%%%%%%%%%%%%%%%% TABLES %%%%%%%%%%%%%%%%%%%%%%%%%%%%%%%%

\begin{deluxetable}{lrrrrrr}
\rotate
\tabletypesize{\normalsize}
\tablecaption{Peak emission measure and temperature values in 8 stellar flares.}
\tablewidth{0pt}
\tablehead{
\colhead{Star}&
\colhead{Peak Emission}&
\colhead{Peak}&
\colhead{Scaling law}&
\colhead{Maximum}&
\colhead{Temperature}&
\colhead{Temperature}\\
\colhead{}&
\colhead{measure}&
\colhead{Temperature}&
\colhead{Temperature}&
\colhead{Temperature}&
\colhead{Difference}&
\colhead{Difference}\\
\colhead{}&
\colhead{$log(EM_p[cm^{-3}$])}&
\colhead{$log(T_p[MK])$}&
\colhead{$log(T_S[MK])$}&
\colhead{$log(T_M[MK])$}&
\colhead{$log(T_S/T_p)$}&
\colhead{$log(T_M/T_p)$}}
\startdata
HR 1099			& 53.8 & 7.61 & 7.67 &    7.69 &    0.06 & 0.08 \\
1EQ1839.6+8002		& 54.0 & 7.88 & 7.71 &    8.03 & $-$0.17 & 0.15 \\
II Peg			& 54.3 & 7.87 & 7.79 & $>$7.87 & $-$0.08 & 0.00 \\
UX Ari			& 55.1 & 7.82 & 7.95 & $>$7.82 & $ $0.13 & 0.00 \\
Algol			& 54.0 & 7.66 & 7.72 &    7.84 &    0.06 & 0.18 \\
AB Dor (Nov 29)		& 54.7 & 8.06 & 7.88 &    8.06 & $-$0.18 & 0.00 \\
AB Dor (Nov 9)		& 54.7 & 7.84 & 7.87 &    7.93 &    0.03 & 0.09 \\
Proxima Centauri	& 51.3 & 7.32 & 7.10 &    7.41 & $-$0.22 & 0.09 \\
			&      &      &      &         &         &      \\
Difference		&      &      &      &         & $-$0.05 & 0.08 \\
Standard deviation	&      &      &	     &         & $\pm$0.13 & $\pm$0.07 \\
\enddata
\end{deluxetable}

\clearpage
%%%%%%%%%%%%%%%%%%%%%%%%%%%%%%%%%%%%%% FIGURE CAPTIONS %%%%%%%%%%%%%%%%%%%%%%%%%%%%%%%%

%\section*{              Figure Captions                                         }

\begin{figure}
\plotone{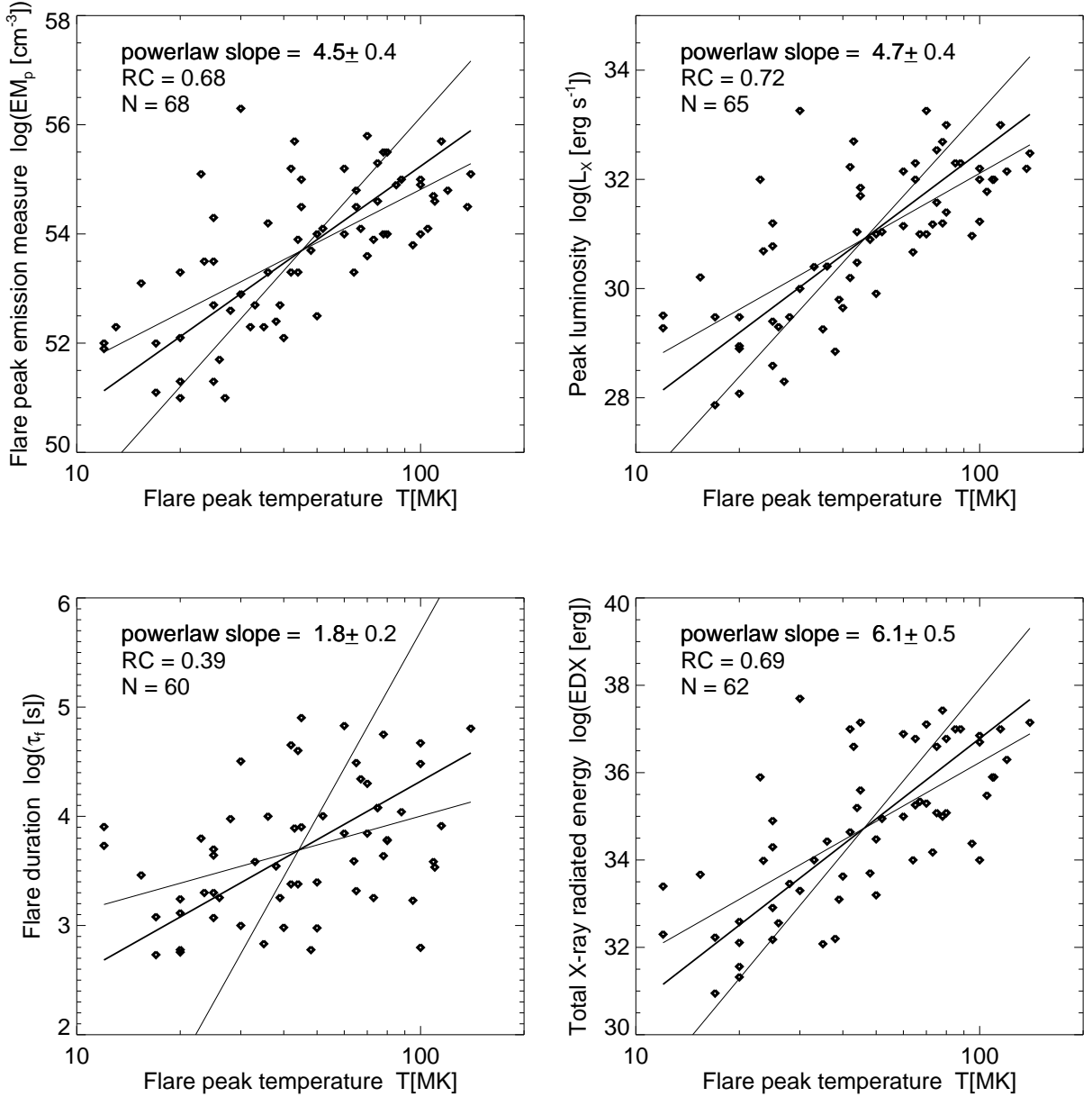}
\caption{Scatterplots of flare peak emission measure $EM_p(T_p)$, peak luminosity
$L_X(T_p)$, flare duration $\tau_f(T_p)$, and total X-ray radiated energy
$E^X(T_p)$ versus the flare peak temperature $T_p$ of 68 stellar flares listed 
in Table 4 of G\"udel (2004). Linear regression fits are shown with thin lines 
[correlating y(x) and x(y)] and with thick lines (ordinary least squares bisector 
method), and the linear regression coefficient (RC) is indicated.}
\end{figure}

\begin{figure}
\vspace*{-10mm}
\plotone{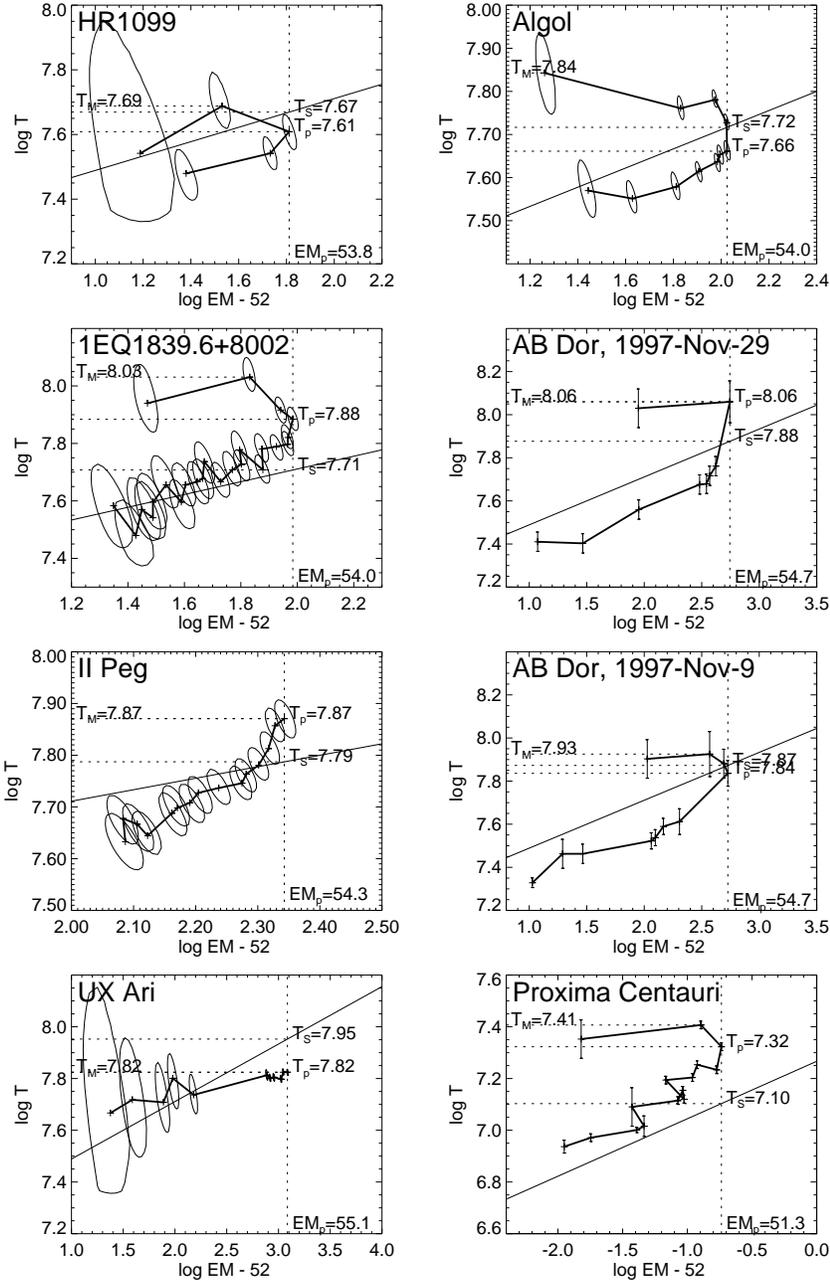}
\caption{Evolutionary phase diagrams of temperature $T(t)$ versus emission measure $EM(t)$
in eight flare events from the following stars: 
HR1099 [or V711 Tau] (Stern 1996), 1EQ1839.6+8002 (Pan et al.~1997), 
II Peg (Doyle et al.~1991), UX Ari (Tsuru et al.~1989), Algol (Stern et al.~1992),
AB Dor (Maggio et al.~2000), and Proxima Centauris (Reale et al.~2003). 
The straight line represents the statistical scaling law
$EM_p \propto T_p^{4.3}$ obtained from G\"udel (2004), which predicts a flare temperature
$T_S$ at the maximum emission measure $EM_p$. For comparison, also the measured flare peak
temperature $T_p$ and the maximum temperature $T_M$ are indicated, listed together in
Table 1. The ellipses shown in 5 cases represent the 90\% confidence intervals in
two-parameter space, while the temperature error bars in the other 3 cases are specified 
in the quoted papers.}
\end{figure}

\begin{figure}
\plotone{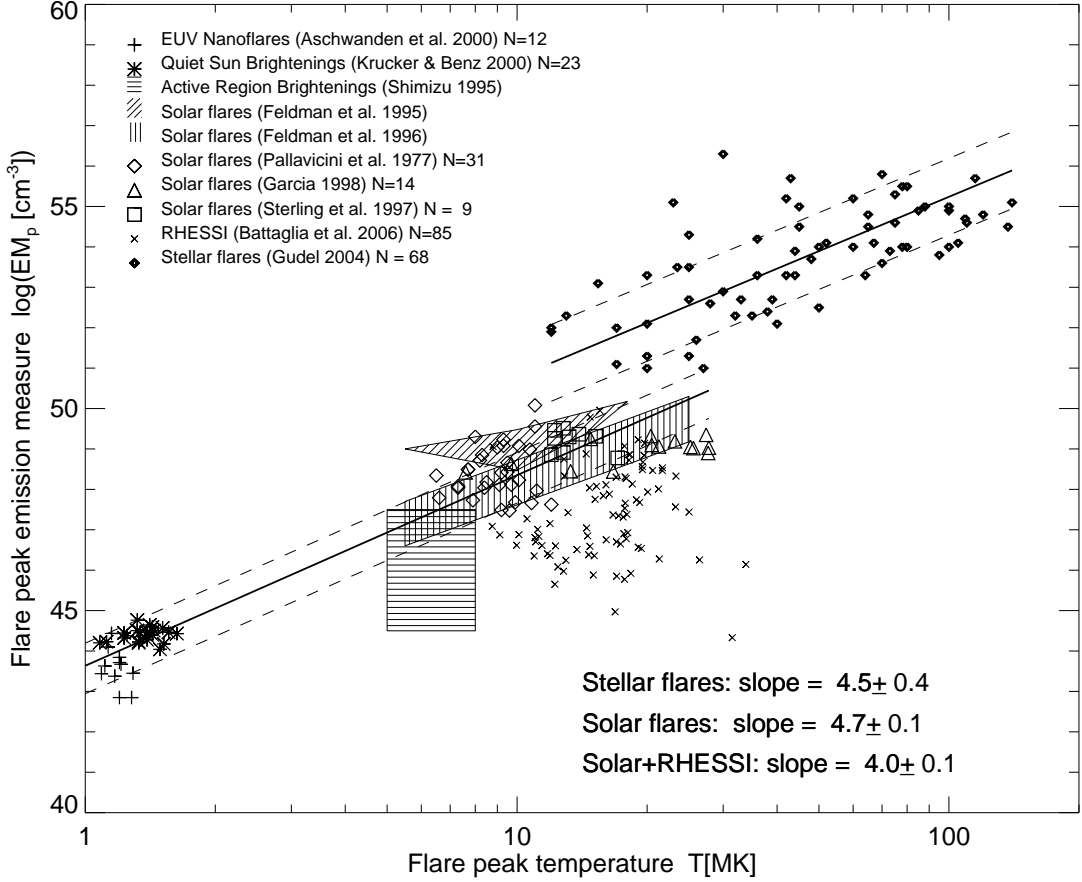}
\caption{Compilation of flare peak emission measure $EM_p$ versus flare peak temperature
$T_p$ measurements in solar and stellar flares. Both the solar and stellar data sets
fit a similar statistical correlation of $EM_p \propto T_p^{4.7}$, but the emissison measures
of stellar flares are about 2 orders of magnitude higher at the same temperature. 
The linear regression fits are shown with thick solid lines, while the 
1$\sigma$-ranges that include 67\% of the datapoints are indicaded with dashed lines.}
\end{figure} 

\begin{figure}
\plotone{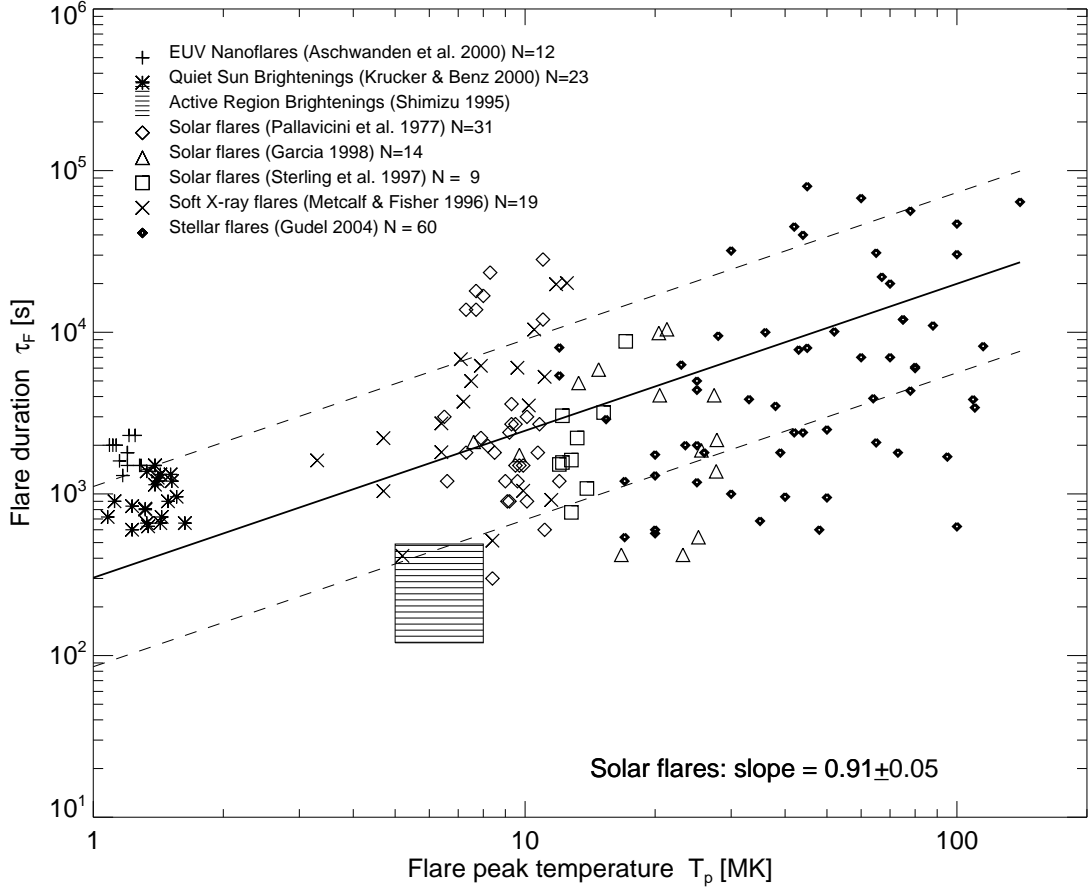}
\caption{Compilation of flare durations $\tau_f$ versus flare peak temperature
$T_p$ measurements in solar and stellar flares. The combined data sets
fit a statistical correlation of $\tau_f \propto T_p^{0.91\pm0.05}$ (thick solid line),
including 67\% of the datapoints within a factor of $\approx 3$ (dashed lines).}
\end{figure}

\begin{figure}
\plotone{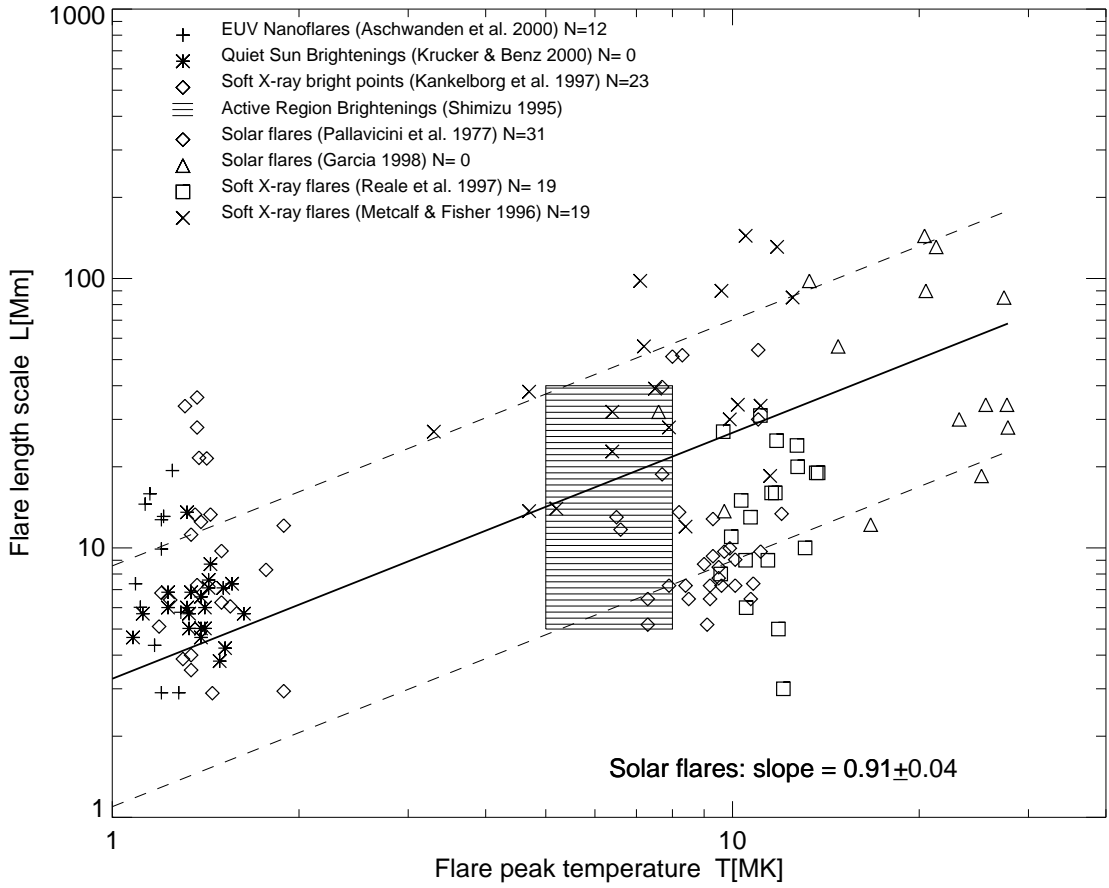}
\caption{Compilation of flare length scales $L$ versus flare peak temperature
$T_p$ measurements in solar and stellar flares. The combined data sets
fit a statistical correlation of $L \propto T_p^{0.91\pm0.04}$ (thick solid line),
including 67\% of the datapoints within a factor of $\approx 3$ (dashed lines).}
\end{figure}

\begin{figure}
\plotone{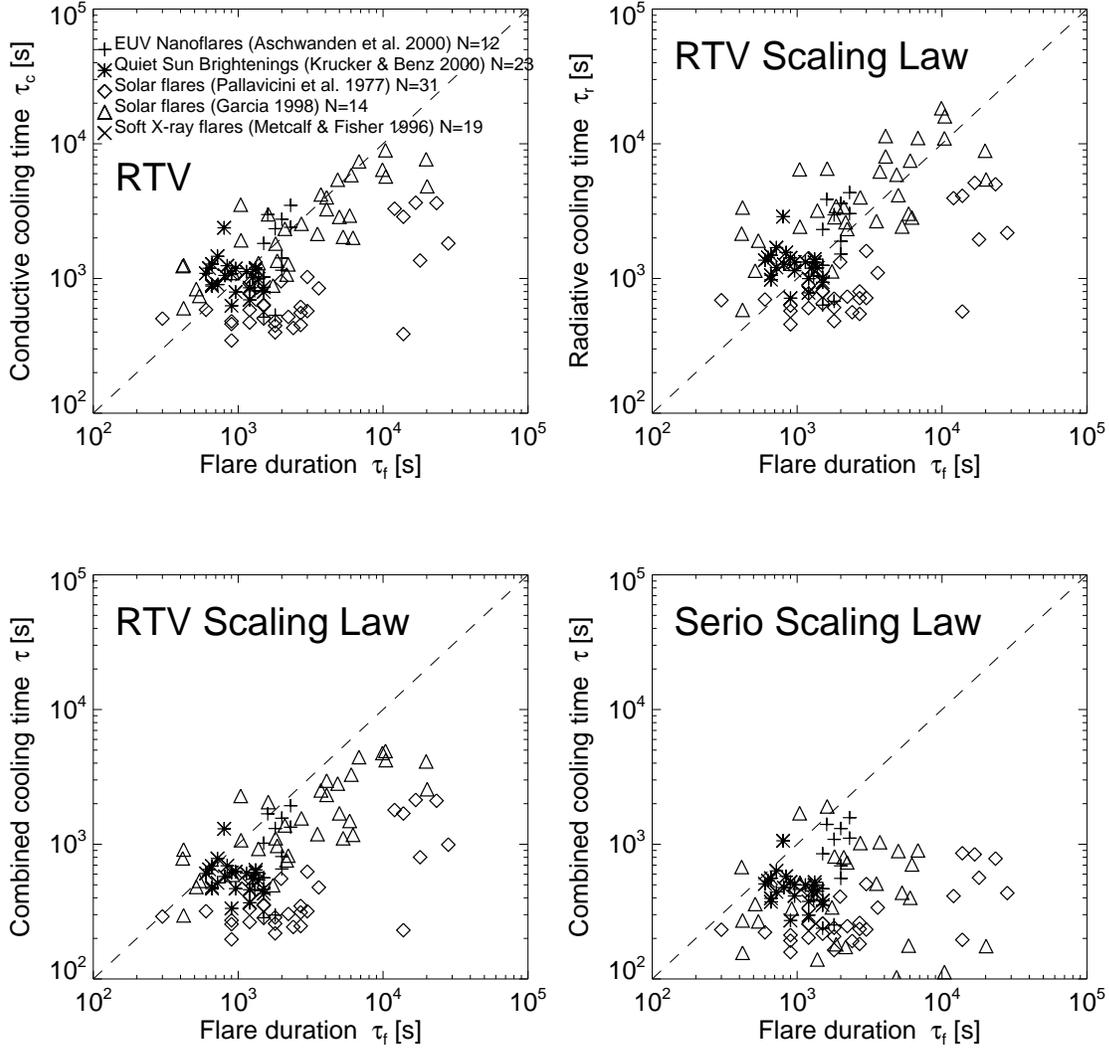}
\caption{Conductive cooling times $\tau_c(L, T_p)$ (top left panel), 
radiative $\tau_c(L, T_p)$ cooling times (top right panel),
combined cooling times $1/\tau=1/\tau_c+1/\tau_r$ for the RTV
scaling law (bottom left panel) and Serio's scaling law (bottom right panel)
are shown as a function of the observed flare duration $\tau_f$, 
calculated with the scaling law using the observed flare peak 
temperatures $T_p$ and flare length scales $L$, and a ratio of $L/s_H=3$ of
the loop half length $L$ to the  heating scale heigth $s_H$, the maximum
limit in Serio's scaling law.}
\end{figure}

\begin{figure}
\plotone{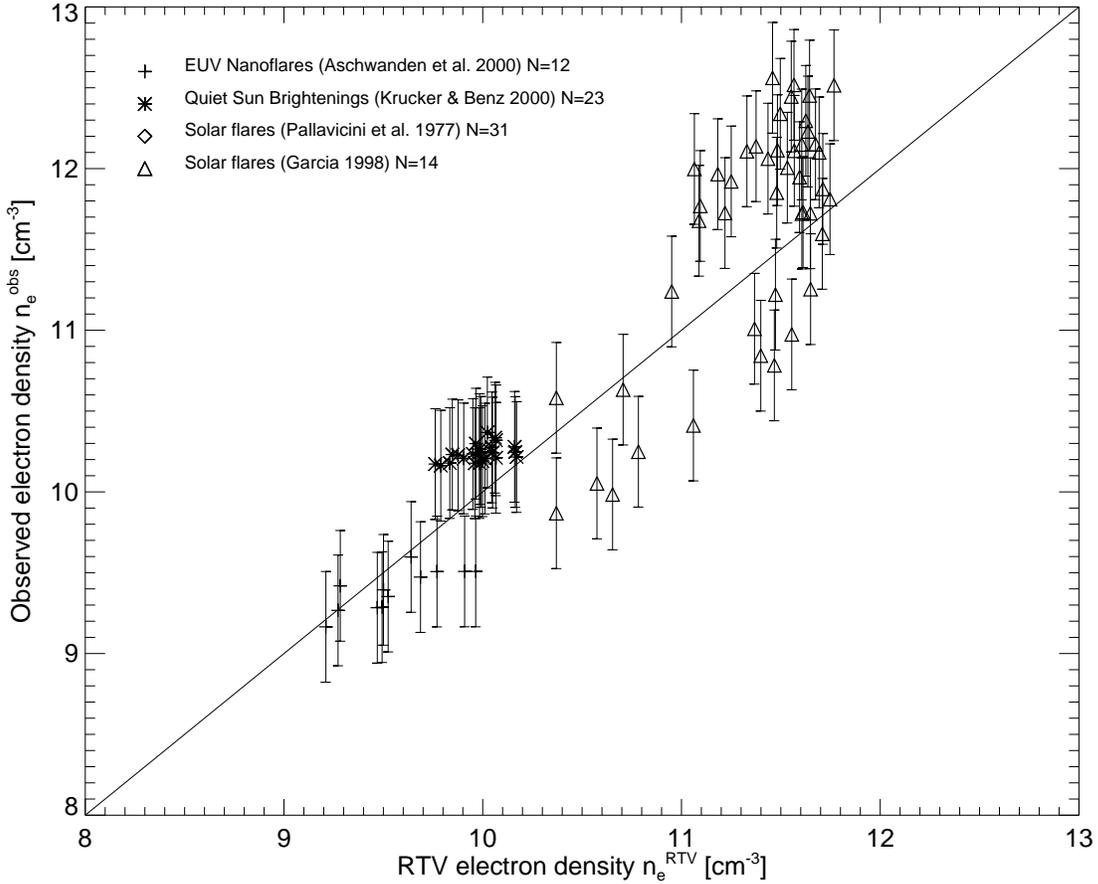}
\caption{Comparison of the electron densities determined in nanoflare and
flare datasets where the flare peak temperature $T_p$, the total volume
emission measure $EM_p$, and the flare length scale $L=A^{1/2}$ has been 
observed, using the theoretical RTV scaling law relationship 
$n_p^{RTV}=c_0 (T_p^2/L)$ (Eq.~50) (x-axis), and the observational 
relationship $n_p^{obs}=\sqrt{EM_p/(q_V L_0^3 (L/L_0)^{D_V})}$ (Eq.~49) (y-axis) 
with $D_V=2.4$ and a volume filling factor range of $q_V=0.03-0.08$ 
at $L_0=25$ Mm (vertical error bars). Both methods predict a range of 
$n_e\approx 10^9-10^{12}$ cm$^{-3}$ and are consistent within less than
an order of magnitude.}
\end{figure}

\begin{figure}
\plotone{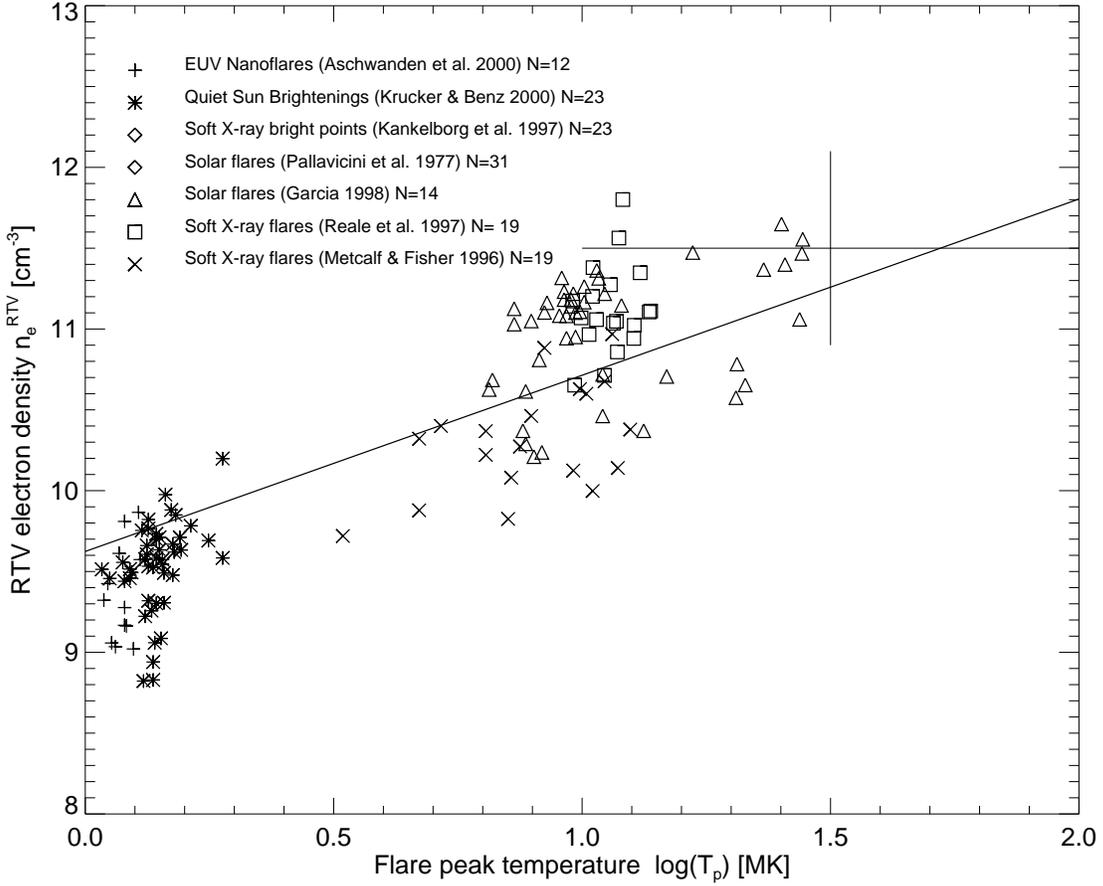}
\caption{Electron densities predicted by the RTV law as a function of the
flare peak temperature, for those datasets with measurements of the flare
peak temperature $T_p$ and length scale $L$ (identical datasets as used in Fig.~5).
The straight line is the statistical prediction $n_p=10^{10.7} (T_p/10$ MK)$^{1.1}$
(Eq.~34). The range of electron densities ($10^{11.5\pm0.6}$ cm$^{-3}$) inferred
for stellar flares according to the compilation of G\"udel (2004) is indicated
with a large cross.}
\end{figure}

\begin{figure}
\plotone{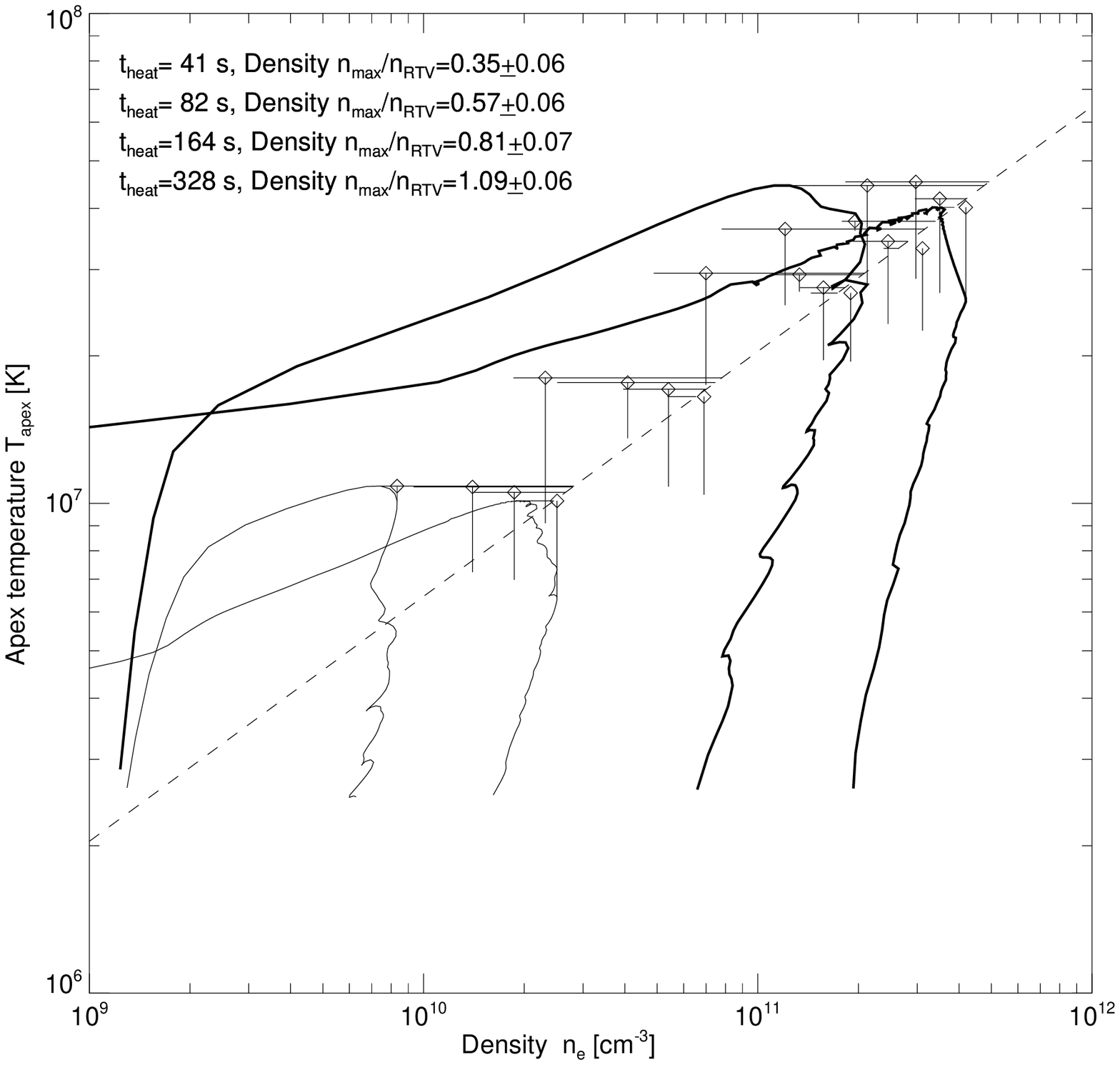}
\caption{Evolutionary curves of $T_{apex}(t)$ versus the electron density
$n_e(t)$ in 20 hydrodynamic simulations of 1D flare loops from Tsiklauri et 
al.~(2004). The parameters are given in Appendix A. All simulations apply to
a flare loop with a length of $L=55$ Mm. We show 4 evolutionary
curves for the extremal heating rates $E_{H0}=0.6$ and $60.0$ erg cm$^{-2}$
s$^{-1}$ and for the extremal heating time scales $t_{heat}=41$ and $329$ s. 
The locations of $T_{max}$ and $n_{max}$ are indicated with diamond symbols. 
The RTV law is indicated with a dashed line. Note the ratios of the maximum flare
densities $n_{max}$ to the values $n_{RTV}$ predicted by the RTV law,
listed at the top of the Figure.}
\end{figure}

\begin{references} %%% REFERENCES
\reference{}{Antiochos, S. K. \& Sturrock, P. A., 1978, ApJ 220, 1137.}
\reference{}{Aschwanden, M. J. 1999, SP 190, 233.}
\reference{}{Aschwanden, M. J., 2004,
	{\sl Physics of the Solar Corona - An Introduction},
	Praxis Publishing Ltd., Chichester UK, and Springer, Berlin} 
\reference{}{Aschwanden, M. J. \& Alexander, D. 2001, Solar Physics, 204, 93}
\reference{}{Aschwanden, M. J. and Nightingale, R. W. 2005, ApJ 633, 499.}
\reference{}{Aschwanden, M. J., Nightingale, R. W., \& Alexander, D. 2000, ApJ 541, 1059.}
\reference{}{Aschwanden, M. J., Nightingale, R. W., \& Boerner, P. 2007, ApJ, (in press).}
\reference{}{Aschwanden, M. J. \& Parnell, C. E. 2002, ApJ 572, 1048.}
\reference{}{Aschwanden, M. J. \& Schrijver, C. J. 2002, ApJS 142, 269.}
\reference{}{Aschwanden, M. J., Schrijver, C. J., and Alexander, D.  2001, ApJ 550, 1036.}
\reference{}{Aschwanden, M. J., Schrijver, C. J., Winebarger, A. R., \& Warren, H. P. 2003, ApJ 588, L49.}
\reference{}{Aschwanden, M. J., Tarbell, T., Nightingale, R., Schrijver, C. J., Title, A., 
	Kankelborg, C. C., Martens, P. C. H., and Warren, H. P. 2000, ApJ 535, 1047.}
\reference{}{Aschwanden, M. J. \& Aschwanden, P. J. 2006, ApJ, (subm).}
\reference{}{Battaglia, M., Grigis, P. C., \& Benz, A. O. 2005, A\&A 439, 737}
\reference{}{Benz, A. O. \& Krucker, S. 1998, Solar Phys. 182, 349}
\reference{}{Culhane, J. L., Phillips, A. T., Inda-Koide, M., Kosugi, T., Fludra, A., Kurokawa, H.,
        Makishima, K., Pike, C. D., Sakao, T., Sakurai, T., Doschek, G. A., \& Bentley, R. D.,
        1994, Solar Phys., 153, 307.}
\reference{}{Doyle, J. G., Kellett, B. J., Byrne, P. B., Avgoloupis, S., Mavridis, L. N.,
	Seiradakis, J. H., Bromage, G. E., Tsuru, T., Makishima, K., Makishima K., 
	McHardy, I. M., 1991, MNRAS 248, 503} 
\reference{}{Feldman, U., Doschek, G. A., Mariska, J. T., \& Brown, C. M. 1995, ApJ 450a, 441.}
\reference{}{Feldman, U., Laming, J. M., \& Doschek, G. A., 1995b, ApJ 451, L79.}
\reference{}{Feldman, U., Doschek, G. A., Behring, W. E., \& Phillips, K. J. H. 1996, ApJ 460, 1034.}
\reference{}{Fisher, G. H., Canfield, R. C., and McClymont, A. N. 1985a, ApJ 281, L79.}
\reference{}{Fisher, G. H., Canfield, R. C., and McClymont, A. N. 1985b, ApJ 289, 414.}
\reference{}{Fisher, G. H., Canfield, R. C., and McClymont, A. N. 1985c, ApJ 289, 425.}
\reference{}{Garcia, H. A. 1998, ApJ 504, 1051.}
\reference{}{G\"udel, M. 2004, AARv 12, 71} 
\reference{}{Hori, K., Yokoyama, T., Kosugi, T., \& Shibata, K. 1997, ApJ 489, 426.}
\reference{}{Hori, K., Yokoyama, T., Kosugi, T., \& Shibata, K. 1998, ApJ 500, 492.}
\reference{}{Isobe, H., Shibata, K., Yokoyama, T., and Imanishi, K. 2003, PASJ 55/5, 967.}
\reference{}{Kankelborg, C. C., Walker, A. B. C. II., \& Hoover, R. B. 1997, ApJ 491, 952}
\reference{}{Krucker, S. \& Benz, A. O. 2000, Sol.Phys. 191, 341}
\reference{}{Lenz, D. D., DeLuca, E. E., Golub, L., Rosner, R., \& Bookbinder, J. A. 1999, ApJ 517, L155.}
\reference{}{MacNeice, P., McWhirter, R. W. P., Spicer, D. S., and Burgess, A. 1984, SP 90, 357.}
\reference{}{Mariska, J. T. and Poland, A. I. 1985, SP 96, 317.}
\reference{}{Metcalf, T. R. \& Fisher, G. H. 1996, ApJ 462, 977.}
\reference{}{Mewe, R, Gronenschild, E. H. B. M., \& van den Oord, G. H. J. 1985,
	A\& AS 62, 197.}
\reference{}{Nagai, F. \& Emslie, A. G. 1984, ApJ 279, 896.}
\reference{}{Ness, J.-U., G\"udel, M., Schmitt, J. H. M. M., Audard, M., and Telleschi, A.
	2004, A\&A 427, 667.}
\reference{}{Pallavicini, R., Serio, S., \& Vaiana, G. S. 1977, ApJ 216, 108.}
\reference{}{Pan, H. C., Jordan, C., Makishima, K., Stern, R. A., Hayashida, K., 
	and Inda-Koide, M. 1997, MNRAS 285, 735.}
\reference{}{Rosner, R., Tucker, W. H., and Vaiana, G. S., 1978, ApJ 220, 643.}
\reference{}{Schmitt, J. H. M. M., Ness, J.-U., and Franco, G. 2003, A\&A 412, 849.}
\reference{}{Shibata, K. \& Yokoyama, A. 1999, ApJ 526, L49.}
\reference{}{Shibata, K. \& Yokoyama, A. 2002, ApJ 577, 422.}
\reference{}{Shimizu, T. 1995, PASJ 47, 251}
\reference{}{Serio, S., Peres, G., Vaiana, G. S., Golub, L., and Rosner, R. 1981, ApJ 243, 288.}
\reference{}{Sterling, A. C., Hudson, H. S., Lemen, J. R., \& Zarro, D. A. 1997, ApJS 110, 115.}
\reference{}{Stern, R. A. 1992, in {\sl Frontiers of X-Ray Astronomy}, (ed. Y. Tanaka \& K. Koyama),
	(Tokyo: Universal Academy Press), 259.}
\reference{}{Stern, R. A. 1996, in {\sl Magnetohydrodynamic Phenomena in the Solar Atmosphere}, 
	IAU Coll. No. 153, Makuhari, Japan, (eds. Y.Uchida, T.Kosugi, and H.S.Hudson), 
	Kluwer Academic Publishers, Dordrecht, p.83}
\reference{}{Stern, R. A., Antiochos, S. K., and Harnden, F. R. Jr. 1986, ApJ 305, 417.}
\reference{}{Stern, R. A., Uchida, Y., Tsuneta, S., \& Nagase, F. 1992, ApJ 400, 321}
\reference{}{Tsiklauri, D., Aschwanden, M. J., Nakariakov, V. M., and Arber, T. D. 2004, A\&A 419, 1149.}
\reference{}{Tsuru, T., Makishima, K., Ohashi, T., Inoue, H., Koyama, K., Turner, M. J. L., 
	Barstow, M. A., McHardy, I. M., Pye, J. P., Tsunemi, H., Kitamoto, S., Taylor, A. R., 
	Nelson, R. F. 1989, PASJ 41/3, 679}
\reference{}{Winebarger, A. R., Warren, H. P., and Mariska, J. T. 2003, ApJ 587, 439.}
\reference{}{Yamamoto, T. T., Shiota, D., Skajiri, T., and Akiyama, S. 2002, ApJ 579, L45.}
\reference{}{Yokoyama, T. and Shibata, K. 1998, ApJ 494, L113.} 
\reference{}{Yokoyama, T. and Shibata, K. 2001, ApJ 549, 1160.}
\end{references}
\end{document}